\documentclass{epuwp}

\usepackage{amsmath}
\usepackage{amssymb}
\usepackage[section]{placeins}
\usepackage{float}

\title{Economic Policy Uncertainty Index for Poland:\\ Data, Construction, and Validation}
\wpsubtitle{}
\wpnumber{01/2026}
\wpshorttitle{EPU Index PL}
\wpshortauthors{Landowska et al.}
\wpdate{October 2026}
\wpversion{}
\wpcutoff{2026-06}
\wpjel{D80, E32, E66, G18, L50}
\wpkeywords{Economic Policy Uncertainty; EPU Index; Poland; News-Based Index; Text Mining; Macroeconomic Uncertainty}
\wpdoi{\href{https://doi.org/10.17605/OSF.IO/EP759}{10.17605/OSF.IO/EP759}}
\wplicense{This paper is licensed under \href{https://creativecommons.org/licenses/by-nd/4.0/}{CC BY-ND 4.0}.}
\wpcitation{Landowska, A.~M., K\l{}opotek, R., \& Zhang, H. (2026).
  \emph{Economic Policy Uncertainty Index for Poland: Data, Construction, and
  Validation}. EPU Index Poland Working Paper No.~01/2026. \url{https://doi.org/10.17605/OSF.IO/EP759}}

\wpauthor{Alina M. Landowska}{1}
\wpauthor{Robert K\l{}opotek}{2}
\wpauthor{He Zhang}{3}
\wpaffiliation{1}{Kozminski University, Warsaw}
\wpaffiliation{2}{Cardinal Stefan Wyszy\'{n}ski University in Warsaw}
\wpaffiliation{3}{The Uncertainty Institute}

\makeatletter
\let\wp@oldbibitem\bibitem
\renewcommand{\bibitem}{\vspace{0.6ex}\wp@oldbibitem}
\makeatother
\begin{document}

\maketitle

%% ---------- Abstract and front matter ---------------------------------
\begin{abstract}
We construct a continuous monthly Economic Policy Uncertainty (EPU) index for Poland spanning January~2000 to June~2026, following the newspaper-based Baker-Bloom-Davis (BBD) methodology. The index is built from six deep-history Polish newspaper titles extracted in-house (\emph{Rzeczpospolita}, \emph{Parkiet}, \emph{Gazeta Wyborcza}, \emph{Dziennik Gazeta Prawna}, \emph{Gazeta Polska Codziennie} and \emph{Puls Biznesu}), scaled by article volume, standardized to unit standard deviation over a strict 25-year fixed baseline (2000--2024, 300 calendar months), averaged across the papers available in each period, and normalized so that the baseline mean equals~100.

Because newspapers enter the panel at different dates, a naive expanding-panel average can jump at each entry point purely from a change in composition rather than from real uncertainty. We correct for this using the BBD Appendix~A splicing procedure: a backward chain-splice from the richest six-paper anchor panel that matches dispersion through a multiplicative scale factor and matches the level through a multiplicative ratio over each overlap window. The resulting index is level-continuous across all newspaper-entry points.

The index exhibits pronounced local increases during selected episodes of economic-policy uncertainty, including the COVID-19 lockdown, the 2022 energy crisis, and the 2023 change of government. The largest local reaction among the selected episodes, however, follows the 2016 Brexit referendum, an externally originated event with significant implications for Poland's economic-policy environment. In month-on-month changes, Poland's EPU is moderately positively correlated with the US (0.26) and global (0.25) EPU indices. Together with the event-based evidence, these relatively modest correlations support a central finding of this study: EPU Index PL captures economic policy uncertainty that is predominantly domestic in orientation and primarily associated with developments affecting Poland's own economic policy, rather than merely reflecting fluctuations in international uncertainty. International shocks remain relevant insofar as they generate uncertainty about economic-policy decisions affecting Poland.
\end{abstract}

\wpfrontmatter
  {}% #1 disclaimer note (not printed)
  {The monthly EPU Index PL series is publicly available on the project website (\url{https://www.epu-index.pl}),GitHub (\url{https://github.com/rakubaku/polish-epu-index}),and the project's OSF repository (\url{https://osf.io/ep759/}). The data are released under the \href{https://creativecommons.org/licenses/by/4.0/}{CC BY 4.0} license.}
  {We are grateful to Professor Steven J. Davis for his valuable methodological discussions and insightful comments on the construction and validation of the EPU Index PL. We also sincerely thank Mediaboard for their essential support in providing the pilot data for the research, without which this study would not have been possible. The remaining historical newspaper archive was extracted in-house by the EPU Index PL team.}
  {Alina M. Landowska, alandowska@kozminski.edu.pl}

%% ---------- Non-technical summary -------------------------------------
\begin{nontechnicalsummary}{Making Economic Policy Uncertainty Measurable in Poland}
\begingroup\small
Economic policy uncertainty is difficult to observe directly, yet understanding how it changes over time is important for central banks, financial institutions, policymakers, and businesses. EPU Index PL provides a systematic measure of uncertainty surrounding economic policy in Poland, drawing on more than 25 years of newspaper coverage.

Based on six Polish newspapers and the internationally established methodology of Baker, Bloom, and Davis (2016), the index tracks economic policy uncertainty from January 2000 to June 2026. Unlike measures of general political attention or financial-market volatility, it focuses specifically on news coverage connecting economic conditions, public policy, and uncertainty.

\begin{keyfindings}
  \item {}\textbf{A consistent measure covering more than 25 years.} EPU Index PL provides a continuous monthly series based on six newspapers: \emph{Rzeczpospolita}, \emph{Parkiet}, \emph{Gazeta Wyborcza}, \emph{Dziennik Gazeta Prawna}, \emph{Gazeta Polska Codziennie} and \emph{Puls Biznesu}. A dedicated adjustment procedure addresses changes in newspaper coverage over time, preventing artificial distortions caused by the addition of new sources.
  \item {}\textbf{Systematic validation of measurement quality.} An independent human review of 600 newspaper articles confirms 70.2\% of articles flagged by the automated method as relevant to economic policy uncertainty. Agreement between reviewers reaches 87.5\% on independently double-coded articles. Additional tests examine sensitivity to editorial orientation and responses to major historical events.
  \item {}\textbf{A Polish perspective not fully captured by global indicators.} Monthly changes in EPU Index PL show modest positive correlations with US EPU (0.26) and Global EPU (0.25). The index therefore provides a distinct national series whose movements are not identical to those of international uncertainty measures.
\end{keyfindings}

\noindent\textbf{Why this matters}

\noindent For \textbf{central banks and policymakers}, EPU Index PL provides a consistent historical indicator for monitoring uncertainty surrounding economic policy in Poland. For \textbf{banks, financial institutions, and researchers}, it offers a documented measure that can be incorporated into subsequent economic and financial analyses.

The contribution of this paper is measurement, not prediction. It establishes how EPU Index PL is constructed, evaluates its reliability and limitations, and provides a foundation for future research into the economic implications of policy uncertainty.

\emph{EPU Index PL measures the intensity of economic policy uncertainty in newspaper coverage. As a transparent, regularly updated, and internationally comparable series, it can be used to track the policy environment alongside conventional indicators, to include policy uncertainty as a variable in forecasting, stress-testing and econometric models, and to compare Poland with the more than 20 countries covered by BBD-consistent EPU indices.}
\par\endgroup
\end{nontechnicalsummary}

%% ---------- Contents --------------------------------------------------
\wpcontents

%% ---------- Main text -------------------------------------------------
\section*{Introduction}
\addcontentsline{toc}{section}{Introduction}

Economic policy uncertainty has become a central concept in macroeconomics and finance because it influences decisions made by households, firms, financial institutions, and policymakers. When the future path of economic policy is uncertain, firms may postpone investment and hiring, households may increase precautionary saving, and financial intermediaries may become more cautious in
their lending decisions. As a result, uncertainty shocks have been linked to lower economic activity, weaker investment, and heightened financial fragility
\citep{bloom2009impact,bloom2014fluctuations,mishkin1999global,phan2020epu}. Understanding these relationships, however, requires reliable measures of policy uncertainty.

The most widely used measure is the newspaper-based Economic Policy Uncertainty (EPU) index developed by \citet{baker2016measuring} (BBD). The EPU index quantifies the share of newspaper articles that simultaneously discuss the economy, public
policy, and uncertainty, providing a transparent, reproducible, and high-frequency indicator that can be consistently implemented across countries.
Since its introduction, EPU indices have become a standard tool for analyzing the effects of policy uncertainty on exchange rates, financial markets, and
macroeconomic outcomes 
\citep{Raunig2021Economic,ZHANG2021102078,abid2020economic,bouoiyour2018measuringresponsegoldprices,phan2020epu}.

The BBD methodology has subsequently been adopted in a growing number of countries, including Germany, Spain, China, Japan, and Belgium
\citep{brandt2021epu,ghirelli2019spainepu,pan2022measuring,arbatli2022policy,algaba2020economic},
as well as in a global, cross-country index, i.e.\ the Global Economic Policy Uncertainty Index (GEPU)
\citep{davis2016gepu}. This growing body of research underscores the importance of constructing nationally representative indices using a transparent and
consistent methodology, as empirical conclusions depend critically on the quality and comparability of the underlying uncertainty measure.

Although Poland has recently attracted growing interest in the measurement of economic policy uncertainty, the literature remains fragmented. Existing studies rely either on a limited set of newspapers, alternative data sources such as Google searches, or shorter historical samples \citep{holda2019newspaper, bartha2022measuring, bialkowski2025local}. Consequently, Poland still lacks a long-run newspaper-based EPU index constructed according to the BBD framework. A key methodological challenge in the Polish context is ensuring comparability over time despite the structural transformation of the newspaper market after 1989. As a post-communist economy, Poland experienced the emergence of new titles, changes in newspaper ownership and readership, and the gradual expansion of digital archives. Without explicit adjustment, changes in the composition of the newspaper panel may introduce artificial level shifts unrelated to actual changes in economic policy uncertainty.

This paper fills this gap by constructing the first long-run newspaper-based Economic Policy Uncertainty Index for Poland fully consistent with the BBD methodology, allowing direct comparability with existing national EPU indices and potential inclusion in future cross-country and global EPU analyses.\footnote{The EPU database provides country-specific EPU indices for 29 countries: Australia, Belgium, Brazil, Canada, Chile, China, Colombia, Croatia, Denmark, France, Germany, Greece, Hong Kong, India, Ireland, Italy, Japan, Mexico, the Netherlands, New Zealand, Nigeria, Pakistan, Russia, Singapore, South Korea, Spain, Sweden, the UK, and the US. The database is available at \url{https://www.policyuncertainty.com}. GEPU is a GDP-weighted aggregate of national EPU indices available through FRED: \url{https://fred.stlouisfed.org/series/GEPUCURRENT}.}
The EPU Index PL is a continuous monthly measure covering January~2000--June~2026, based on six Polish newspapers with long historical archives.

Our contribution is fourfold. First, we construct the longest continuous newspaper-based measure of domestic economic policy uncertainty available for
Poland, normalized using a strict twenty-five-year fixed baseline. Second, we maintain comparability despite changes in newspaper composition by implementing the backward chain-splice procedure proposed by BBD and provide empirical evidence on its performance. Third, we validate the classification procedure through manual annotation, reporting precision, inter-annotator agreement, and a detailed taxonomy of false positives. Fourth, we examine whether the index is robust to differences in newspapers' editorial orientation.

\section{Data}

The Polish newspaper market reflects the gradual evolution of the media landscape after 1989. The dataset is based on in-house extractors that, for each title and each day, produce a count of EPU-matching articles ($C$) and a count of total articles ($N$). Table~\ref{tab:newspaper_history} presents the historical background, digital archive availability, and panel entry points of the six newspaper titles used in the EPU Index PL.

\begin{table}[!t]
\centering
\caption{The six newspaper titles included in the EPU Index PL: historical background, data coverage, and panel entry}
\label{tab:newspaper_history}
\small
\begin{tabular}{p{3.2cm}p{2cm}p{2.5cm}p{6cm}}
\toprule
\textbf{Newspaper} & \textbf{Founded} & \textbf{Data coverage} & \textbf{Historical background} \\
\midrule

Rzeczpospolita
& 1920 (1944--50)
& 2000--2026
& Re-established in 1982 during the communist period; it provides the longest continuous newspaper archive available in the sample. \\

Gazeta Wyborcza
& 1989
& 2009--2026
& Founded during Poland's democratic transition. \\

Parkiet
& 1994
& 2000--2026
& Poland's first nationwide financial daily. \\

Dziennik Gazeta Prawna
& 1994
& 2009--2026
& \emph{Gazeta Prawna} launched as a weekly in 1994, became a daily newspaper in 2001, and merged with \emph{Dziennik Polska--Europa--Świat} in 2009 to form \emph{Dziennik Gazeta Prawna}. \\

Puls Biznesu
& 1997
& 2014--2026
& Nationwide business daily. \\

Gazeta Polska Codziennie
& 2011
& 2011--2026
& National daily with extensive political and economic coverage. \\

\bottomrule
\end{tabular}

\begin{flushleft}
\footnotesize
\textit{Notes:} Several newspapers existed before 2000; however, complete and consistent digital archives are not available for their earliest years. Therefore, the EPU Index PL uses the earliest period for which digitized newspaper archives allow consistent article-level extraction and cross-title comparison. Data coverage refers to the period used in the index, not
necessarily the year of the newspaper's establishment. During the construction phase, additional newspaper titles were considered. \emph{Gazeta Podatkowa} was
excluded from the final panel because its inclusion generated a composition-driven break unrelated to changes in economic policy uncertainty. The final
specification uses six titles with stable and comparable digital archive coverage.
\end{flushleft}
\end{table}

The merged daily panel underlying the index contains 9{,}123 dated observations. Coverage is not uniform across titles: the number of days carrying a non-empty article count ($N>0$) reflects each title's own archive depth and publication schedule, and ranges from 7{,}801 days for \emph{Rzeczpospolita}, the longest continuous archive in the panel, to 3{,}052 for \emph{Puls Biznesu}, which enters in 2014. The remaining titles fall between these bounds: \emph{Parkiet} (7{,}352), \emph{Gazeta Wyborcza} (6{,}296), \emph{Dziennik Gazeta Prawna} (4{,}222) and \emph{Gazeta Polska Codziennie} (4{,}109). These differences are accommodated by the chain-splicing procedure of  Section~\ref{sec:splice_robustness}, which adjusts for panel composition at each entry point rather than assuming a balanced panel.

\newpage
\section{Methodology}
\label{ch:methodology}

\subsection{The Economy-Policy-Uncertainty Framework}
\label{sec:epu_framework}

Following the BBD Economy--Policy--Uncertainty (E/P/U) framework, an article is classified as EPU-related if it contains at least one keyword from each of the three categories: Economy, Policy, and Uncertainty. The original BBD dictionary is adapted to the Polish language and institutional environment. Table~\ref{tab:epu_dictionary} summarizes the resulting mapping used in the EPU Index PL.

\begin{table}[!htbp]
\centering
\caption{E/P/U dictionary: BBD framework and Polish adaptation}
\label{tab:epu_dictionary}
\small
\renewcommand{\arraystretch}{1.15}
\begin{tabular}{p{0.20\textwidth} p{0.28\textwidth} p{0.42\textwidth}}
\toprule
\textbf{Category} & \textbf{BBD framework} & \textbf{Polish adaptation (EPU Index PL)} \\
\midrule
\textbf{E: Economy}
& economy
& gospodar* / koniunktur* \\
& economic
& \\
& economics
& ekonomi* / makroekonomi* / mikroekonomi* \\
\midrule
\textbf{P: Policy}
& policy
& polit* \\
& spending
& wydatk* \\
& budget
& budżet* \\
& tax
& podatek / podatk* \\
& deficit
& deficyt* \\
& regulation
& regulac* \\
& legislation
& ustaw* / legislac* / rozporządz* / dyrektyw* \textit{(adapted)} \\
& Congress / legislature
& sejm* / sena* / parlament* / marszał* \textit{(adapted)} \\
& government
& rząd* \textit{(adapted)} \\
& finance ministry / minister
& minist* \textit{(adapted)} \\
& white house
& prezyden* \textit{(adapted)} \\
& --
& UE / Uni* Europejsk* / Bruksel* / Komisj* Europejsk* \\
&
& \quad / PE / europarlament* \textit{(v5.1 only, adapted)} \\
& Federal Reserve / central bank
& NBP / Narodow* Bank* Polsk*  / bank* centraln* \\
&
& \quad / RPP / Rad* Polity* Pienięż* \textit{(adapted)} \\
&
& \quad / Europejsk* Bank* Centraln* \textit{(adapted)} \\
&
& \quad / EBC / KE \textit{(v5.1 only, adapted)} \\
\midrule
\textbf{U: Uncertainty}
& uncertain / uncertainty
& niepewność / niepewności / niepewną / niepewny / niepewna / niepewne / niepewnym / niepewnymi  / niepewnych / niepewnie / niepewnego / niepewnej / niepewnemu / niepewni / brak* pewności \textit{(adapted)} \\
\bottomrule
\end{tabular}
\begin{flushleft}
\footnotesize
\textit{Notes:} The P term set follows the common core used by BBD in their country-level indices for the UK, Canada, France, Italy, and Spain (spending, policy, deficit, budget, tax, regulation, and the national central bank; cf.\ their Appendix~A). Congress/legislature, White House, and legislation are drawn from the BBD US term set and mapped to their Polish institutional counterparts; government and finance ministry/minister terms appear in BBD's term sets for South Korea, China, and India. European Union-related terms are included as a country-specific institutional adaptation, reflecting the importance of EU-level regulation for Polish economic policy uncertainty. For morphologically rich languages, BBD themselves enumerate inflected forms explicitly rather than relying on wildcards (e.g.\ their French and Italian term sets); we follow this precedent for the U category, listing inflected forms of the root \emph{niepewn-} directly. We additionally include the synonymous analytic construction \emph{brak pewności} (``lack of certainty''), which captures the same meaning through a different lexical structure not reachable by the wildcard.
\end{flushleft}
\end{table}

The policy dictionary was developed through six successive versions (v1--v5), each evaluated against the 600-article human audit sample on the gross error criterion (Section~\ref{sec:human_audit}); v5 attained the lowest gross error among the six and was adopted as the base term set. As shown in Table~\ref{tab:dictionary_versions}, successive versions reduced linguistically identifiable false positives while preserving conceptual comparability with the original BBD methodology. Consequently, the share of articles classified as EPU decreased from 8.930\% to 1.176\% under v5. The final specification proved highly robust: correcting the exclusion-rule implementation (v4.1$\rightarrow$v4.2) recovered only 82 additional articles out of 1{,}986{,}139 (0.0041\% of the corpus), without changing the dictionary itself. The published index uses dictionary version~v5.1, which adds four European-institution terms to v5's Policy category (\emph{PE}, \emph{europarlament*}, \emph{EBC}, \emph{KE} -- see Table~\ref{tab:dictionary_diff}, Appendix~\ref{app:dictionary}), spliced under the multiplicative convention of Section~\ref{sec:additive_vs_multiplicative} (series \texttt{epu\_strict\_v51\_mult}); this is the main and final version of the EPU Index PL. The 600-article audit sample itself was drawn and coded against v5 (Section~\ref{sec:human_audit}). Because v5.1 only adds terms to v5's Policy category and removes none, every article classified EPU$=$1 under v5 is also EPU$=$1 under v5.1 (strict containment, verified directly on the same 2000-01--2026-06 index window and six-title panel used for $N^{+}$ in Section~\ref{sec:audit_sample_design}: 0 of 23{,}357 v5-positive articles are v5.1-negative); v5.1 additionally flags 47 articles not caught by v5, out of 1{,}986{,}139 articles in that window and panel (0.0024\% of the corpus). We further verified this containment on the audit sample itself: all 480 flagged articles drawn and coded under v5 were checked against the live v5.1 classification, and none of them is among the 47 articles added by v5.1. The 600-article audit sample's confusion matrix (see Table~\ref{tab:confusion_matrix}) is therefore identical whether evaluated under v5 or v5.1, and the gross-error comparison in Table~\ref{tab:gross_error_versions} did not need to be separately recomputed for v5.1.

\begin{table}[!htbp]
\centering
\caption{Iterative refinement of the EPU Index PL keyword dictionary}
\label{tab:dictionary_versions}
\small
\setlength{\tabcolsep}{4pt}
\begin{tabular}{llc}
\toprule
\textbf{Version} & \textbf{Main methodological change} & \textbf{\shortstack{Share of articles\\classified as EPU}} \\
\midrule
v1 & Legacy implementation (extended keyword lists) & 8.930\% \\
v2 & Removal of obvious false positives & 3.948\% \\
v3 & Stem-based implementation & 3.339\% \\
v4.1 & Introduction of exclusion rules & 1.090\% \\
v4.2 & Corrected exclusion-rule implementation (+82 articles) & 1.094\% \\
v5 & Base BBD-consistent dictionary (lowest gross error among v1--v5) & 1.176\% \\
v5.1 & Published version: v5 + PE, europarlament*, EBC, KE (Policy) & 1.178\% \\
\bottomrule
\end{tabular}

\vspace{0.2cm}
\begin{flushleft}
\footnotesize
\textit{Note:} The reported values denote the percentage of newspaper articles classified as EPU within the 2000-01--2026-06 index window and the six-title panel (1,986,139 articles), the same scope as $N^{+}$ in Section~\ref{sec:audit_sample_design}. The increase from v4.1 to v4.2 reflects only a correction to the implementation of the exclusion rules, recovering 82 valid EPU articles while leaving the underlying keyword dictionary unchanged.
\end{flushleft}
\end{table}

\begin{figure}[!htbp]
\centering
\includegraphics[width=0.95\textwidth]{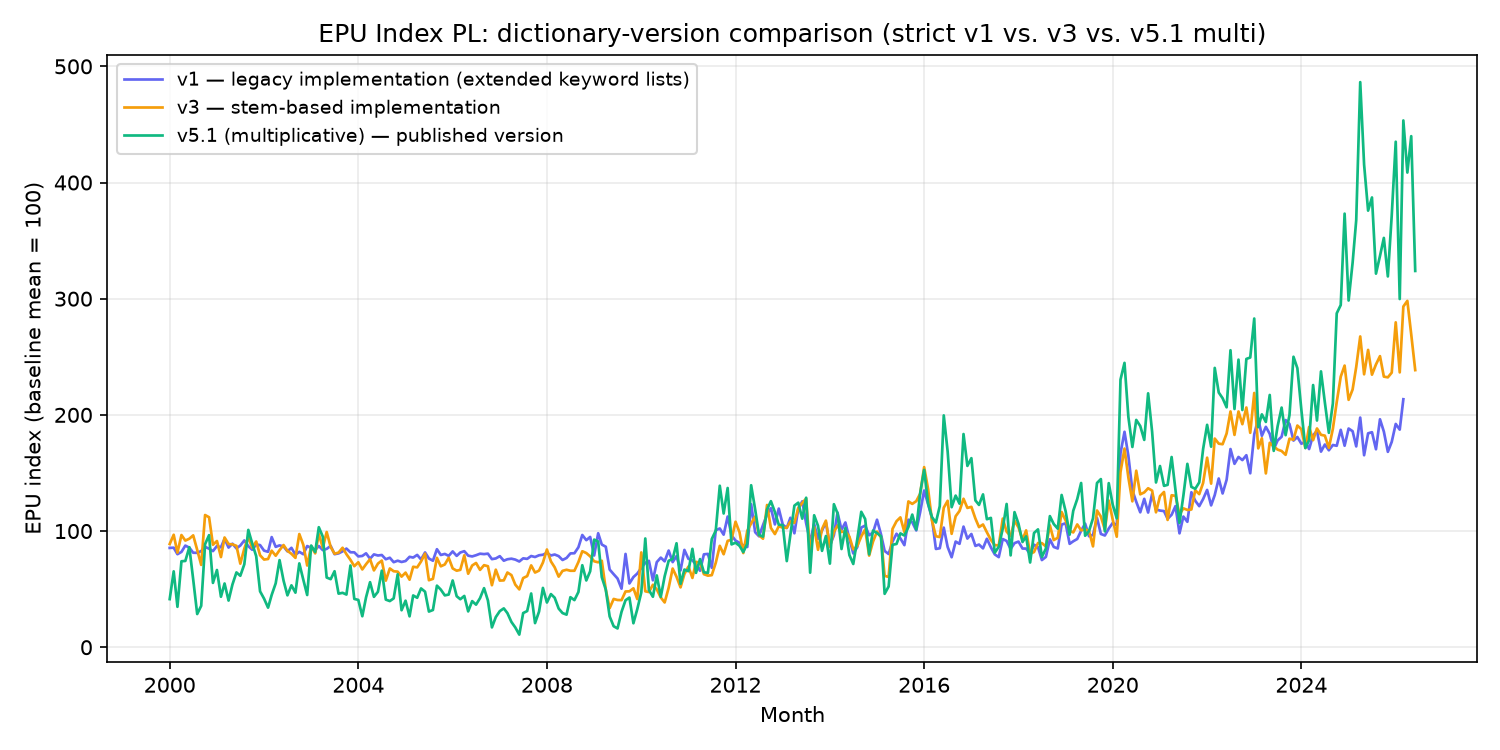}
\caption{EPU Index PL: dictionary-version comparison (strict v1 vs.\ v3 vs.\ v5.1 multiplicative)}
\label{fig:dictionary_versions}
\wpsource{Authors' calculations.}
\end{figure}

Figure~\ref{fig:dictionary_versions} compares the index obtained under dictionary versions v1, v3 and v5.1 (multiplicative splice).

\subsection{Index construction}
\label{sec:index_construction}

Following the procedure set out in Section~2.1 of \citet{baker2016measuring}, the monthly index is computed in five steps. Let $C_{it}$ denote the number of articles in newspaper $i$ and month $t$ that satisfy the E$\wedge$P$\wedge$U criterion, and $N_{it}$ the total number of articles published by the same newspaper in the same month. First, for each newspaper, the raw count is scaled by article volume:
\begin{equation}
X_{it}=\frac{C_{it}}{N_{it}}.
\label{eq:scaling}
\end{equation}
Second, each newspaper-level series is standardized to unit standard deviation over the fixed 2000--2024 baseline,
\begin{equation}
Y_{it}=\frac{X_{it}}{\sigma_i},
\label{eq:standardization}
\end{equation}
where $\sigma_i$ is the standard deviation of $X_{it}$ over the baseline months in which newspaper $i$ is available. Third, the standardized series are averaged with equal weights across the $n_t$ newspapers available in month $t$:
\begin{equation}
Z_t=\frac{1}{n_t}\sum_{i=1}^{n_t} Y_{it}.
\label{eq:panel_average}
\end{equation}
Fourth, the mean $M$ of $Z_t$ over the baseline is computed. Fifth, $Z_t$ is rescaled so that the index has a baseline mean of 100:
\begin{equation}
\mathrm{EPU}_t=\frac{100}{M}\,Z_t.
\label{eq:normalization}
\end{equation}
Because newspapers enter the panel at different dates, the averaging in Equation~\eqref{eq:panel_average} is combined with the splicing procedure described in Section~\ref{sec:splice_robustness}.

\subsection{Adjusting for Newspaper Composition: BBD Appendix-A Splicing}
\label{sec:splice_robustness}
The final EPU Index PL is constructed using the backward chain-splice procedure described in Appendix~A of \citet{baker2016measuring}, with the six-newspaper panel used as the anchor specification. The historical panel begins with a Rzeczpospolita--Parkiet segment (2000--2008), followed by successive newspaper additions. The splice chain  (P2$\rightarrow$P4, P4$\rightarrow$P5, P5$\rightarrow$P6) and adjustment parameters are reported in Table~\ref{tab:splice}.

\begin{table}[!htbp]
\centering
\caption{Monthly splice parameters (backward chain from the six-paper anchor), computed on dictionary version v5.1}
\label{tab:splice}
\footnotesize
\setlength{\tabcolsep}{3.5pt}
\begin{tabular}{lccccc}
\toprule
Splice step & Entry event & \shortstack{Overlap\\(m.)} & Corr. & \shortstack{$\sigma$-scale / shift\\(additive, BBD)} & \shortstack{Ratio\\(multiplicative, adopted)} \\
\midrule
P5 $\to$ P6 & + Puls Biznesu (2014)      & 150 & 0.9930 & 0.9747 / $+0.0164$ & 1.0077 \\
P4 $\to$ P5 & + GPC (2011)               & 178 & 0.9802 & 1.0781 / $-0.3254$ & 0.8538 \\
P2 $\to$ P4 & + Wyborcza + DGP (2009)    & 210 & 0.9467 & 0.8723 / $+0.1880$ & 1.3217 \\
\bottomrule
\end{tabular}
\vspace{3pt}
\begin{flushleft}
\footnotesize
\textit{Notes:} The additive shift column follows BBD (2016) Appendix A exactly; the multiplicative ratio column reports the convention actually used to construct the published series (Section~\ref{sec:additive_vs_multiplicative}). All parameters are computed on dictionary version v5.1 (\texttt{strict\_v5\_1}), the same data underlying the published \texttt{epu\_strict\_v51\_mult} series.
\end{flushleft}
\wpsource{Authors' calculations.}
\end{table}

To illustrate the mechanics of the procedure, we describe the P2$\to$P4 step (\emph{Rzeczpospolita} and \emph{Parkiet} joined by \emph{Gazeta Wyborcza} and \emph{Dziennik Gazeta Prawna} in 2009) under the additive convention in detail. Panel P2's standardized series $Z^{P2}_t$ is compared against panel P4's already-adjusted series $Z^{P4}_t$ over their overlap window (2009-01 to 2026-06, $n=210$ months, $\text{corr}=0.947$). The variance-matching scale and the level-matching shift are
\begin{equation}
\text{scale}=\frac{\sigma(Z^{P4})}{\sigma(Z^{P2})}=0.872,\qquad
\text{shift}=\overline{Z^{P4}}-\text{scale}\cdot\overline{Z^{P2}}=+0.188,
\label{eq:splice_params}
\end{equation}
so the adjusted P2 series under the additive (BBD) convention is
\begin{equation}
\widehat{Z}^{P2}_t = 0.872\cdot Z^{P2}_t + 0.188.
\label{eq:splice_additive}
\end{equation}

Under the multiplicative convention actually adopted for the published series (Section~\ref{sec:additive_vs_multiplicative}), the P2$\to$P4 variance-matching scale is \emph{not} the same number, because the chain is applied sequentially from P6 backward and the P4$\to$P5 step's multiplicative ratio rescales $Z^{P4}$'s standard deviation (whereas the additive shift at that same step leaves it unchanged). Only the first link in the chain, P5$\to$P6, has an identical scale under both conventions; from P4$\to$P5 onward the two conventions compound different adjusted standard deviations into every subsequent step's scale. For P2$\to$P4 specifically, the multiplicative scale is $0.751$ (versus $0.872$ additive), and it is combined with a level-matching ratio of $1.322$, giving
\begin{equation}
\widehat{Z}^{P2}_t = 0.751\cdot 1.322\cdot Z^{P2}_t = 0.992\cdot Z^{P2}_t,
\label{eq:splice_multiplicative}
\end{equation}
the series that actually feeds the final spliced output for 2000--2008.

A potential limitation of this procedure arises because the additive shift is fitted on the data-rich 2009+ overlap and then applied unchanged to the sparser 2000--2008 own segment. As a result, a large event-driven spike in the two-paper series occurring \emph{before} the overlap window may be under-represented in the final spliced series. The 2008--2009 Global Financial Crisis (GFC) illustrates this point: a large increase recorded during the GFC by the two titles spanning the full sample (Rzeczpospolita and Parkiet) may not be fully reflected in the spliced series.

To examine this issue, Table~\ref{tab:davis_gfc} compares the GFC-window (2008-01 to 2009-12) jump in EPU Index PL level, relative to the 2007 average, across three constructions: (i) the unspliced two-paper (Rzeczpospolita+Parkiet) series, used as an unadjusted benchmark; (ii) the official spliced series under the original additive Appendix-A convention; and (iii) the spliced series under a multiplicative variant of the same procedure (Section~\ref{sec:additive_vs_multiplicative}), which matches the level via a ratio rather than an additive shift.

\begin{table}[!htbp]
\centering
\caption{GFC-window level jump (2007 average to 2008--2009 peak), by construction}
\label{tab:davis_gfc}
\small
\begin{tabular}{lccc}
\toprule
Construction & 2007 average & 2008--09 peak & Jump \\
\midrule
Two-paper baseline (RP+Parkiet), unspliced & 47.6 & 119.2 & $+71.6$ \\
Official spliced, additive (robustness variant) & 61.6 & 106.6 & $+45.1$ \\
Official spliced, multiplicative & 46.9 & 114.1 & $+67.2$ \\
\bottomrule
\end{tabular}
\begin{flushleft}
\footnotesize
\textit{Notes:} Two-paper baseline uses panel P2 (Rzeczpospolita+Parkiet), which spans the full 2000--2026 sample with no entry gap and therefore requires no chain-splicing; it is a direct, unscaled read of the raw signal. The multiplicative row is the main-text specification (series \texttt{epu\_strict\_v51\_mult}); the additive row is reported as a labeled robustness variant only.
\end{flushleft}
\end{table}

The additive-splice jump of 45.1 points is materially smaller than the 71.6-point jump in the unspliced two-paper series, indicating that the additive procedure attenuates the true GFC-period signal. The multiplicative variant, introduced to address a separate problem (spurious negative index values in narrow-dictionary term-set variants; Section~\ref{sec:additive_vs_multiplicative}), also resolves this attenuation: its GFC jump of 67.2 points is of the same order of magnitude and direction as the unspliced reference. Both issues -- attenuation of pre-overlap events and negative post-splice values -- share a common source: an \emph{additive} shift fitted on a variance/level regime that differs from the region to which it is applied.

As a complementary robustness check, we consider delaying a newspaper's effective entry into the spliced series from its first overlap date to a later date characterized by a calmer economic and policy environment, specifically by moving the P4 (Gazeta Wyborcza + DGP) entry from 2009-01 to 2011-01. This leaves an eight-month window (2011-01 to 2011-08) in which P4 is the active panel on its own, immediately followed by the already-scheduled entry of P5 (adding Gazeta Polska Codziennie) on 2011-09-01, GPC's genuine first-issue date; the two entries are close in time but not simultaneous. The delayed entry is implemented as an optional setting of the splice procedure, and the comparison is repeated under this specification (see Table~\ref{tab:davis_delayed_entry}).

\begin{table}[!htbp]
\centering
\caption{GFC-window jump under delayed panel entry (2009 vs.\ 2011), by splice method}
\label{tab:davis_delayed_entry}
\small
\begin{tabular}{lc}
\toprule
Construction & Jump \\
\midrule
Two-paper baseline (reference) & $+71.6$ \\
Additive, P4 entry 2009-01 (original) & $+45.1$ \\
Additive, P4 entry delayed to 2011-01 & $+52.1$ \\
Multiplicative, P4 entry 2009-01 & $+67.2$ \\
Multiplicative, P4 entry delayed to 2011-01 & $+71.2$ \\
\bottomrule
\end{tabular}
\end{table}

Delaying entry helps only marginally under the additive convention (45.1$\to$52.1). Combined with the multiplicative convention, however, delayed entry recovers a jump of 71.2 points, matching the two-paper reference (71.6) to within 0.4 points. The multiplicative convention alone, at the standard 2009 entry date, already recovers most of this gain (67.2 of 71.6 points, Table~\ref{tab:davis_gfc}), and is the specification used throughout the remainder of this paper (series \texttt{epu\_strict\_v51\_mult}, Table~\ref{tab:splice}). We therefore adopt multiplicative splicing at the standard entry dates as our main specification, and report both the additive convention and the delayed-entry (2011) variant as labeled robustness checks (see Table~\ref{tab:davis_delayed_entry}).

\subsubsection{Additive vs.\ multiplicative level-matching}
\label{sec:additive_vs_multiplicative}

The BBD Appendix-A procedure matches the level over the overlap window by an additive shift. Because that shift is fitted on the overlap and then applied unchanged to a different (typically sparser, narrower-dictionary) regime, it can push the adjusted series below zero when the source segment's own level is close to zero and the fitted shift is negative relative to it. This occurs in practice for each of the narrow-dictionary term-set variants considered in this study -- \emph{strict\_v4.1}, \emph{strict\_v4.2}, \emph{strict\_v5} (the base term set of Table~\ref{tab:dictionary_versions}) and \emph{strict\_v5.1} (v5 plus four European-institution terms, adopted for all main-text results) -- and not only for the narrowest of the four: under the additive convention, all four monthly series have exactly 3 negative months, with minima ranging from $-7.82$ (v5) to $-9.57$ (v4.2), values that are not interpretable for an index constructed as a non-negative, volume-scaled article frequency.

We therefore compute every one of these four EPU versions under a \emph{multiplicative} level-matching convention: the level is matched by a ratio rather than a shift, $\widehat{Z}^{src}_t=\mathrm{scale}\cdot\mathrm{ratio}\cdot Z^{src}_t$, so that a non-negative source series is guaranteed to remain non-negative after splicing. Applying this to all four versions eliminates every negative value in every case: minima rise to between $+10.85$ (v5.1) and $+11.26$ (v4.1), on the identical 318 monthly observations per version. Table~\ref{tab:additive_vs_multiplicative} summarizes both improvements -- the level-continuity gain documented here and the GFC-signal-attenuation gain documented in Tables~\ref{tab:davis_gfc}--\ref{tab:davis_delayed_entry} -- since both trace to the same additive-shift mechanism.

\begin{table}[!htbp]
\centering
\caption{Additive vs.\ multiplicative splicing: monthly series, all four narrow-dictionary versions}
\label{tab:additive_vs_multiplicative}
\small
\begin{tabular}{lcccccc}
\toprule
& \multicolumn{3}{c}{Additive (original)} & \multicolumn{3}{c}{Multiplicative (adopted)} \\
Version & Min & Max & Neg.\ months & Min & Max & Neg.\ months \\
\midrule
v4.1 & $-9.49$  & $488.30$ & 3 & $+11.26$ & $493.03$ & 0 \\
v4.2 & $-9.57$  & $485.35$ & 3 & $+11.21$ & $490.45$ & 0 \\
v5   & $-7.82$  & $480.04$ & 3 & $+10.88$ & $487.79$ & 0 \\
v5.1 & $-7.88$  & $478.59$ & 3 & $+10.85$ & $486.42$ & 0 \\
\bottomrule
\end{tabular}
\begin{flushleft}
\footnotesize
\textit{Notes:} All four narrow-dictionary term-set variants (318 monthly observations each), computed from the identical underlying article counts per version; only the splice-procedure level-matching step differs. v5.1 (v5 plus four European-institution terms) is the version adopted for the main-text results.
\end{flushleft}
\end{table}

\begin{figure}[H]
\centering
\includegraphics[width=0.95\textwidth]{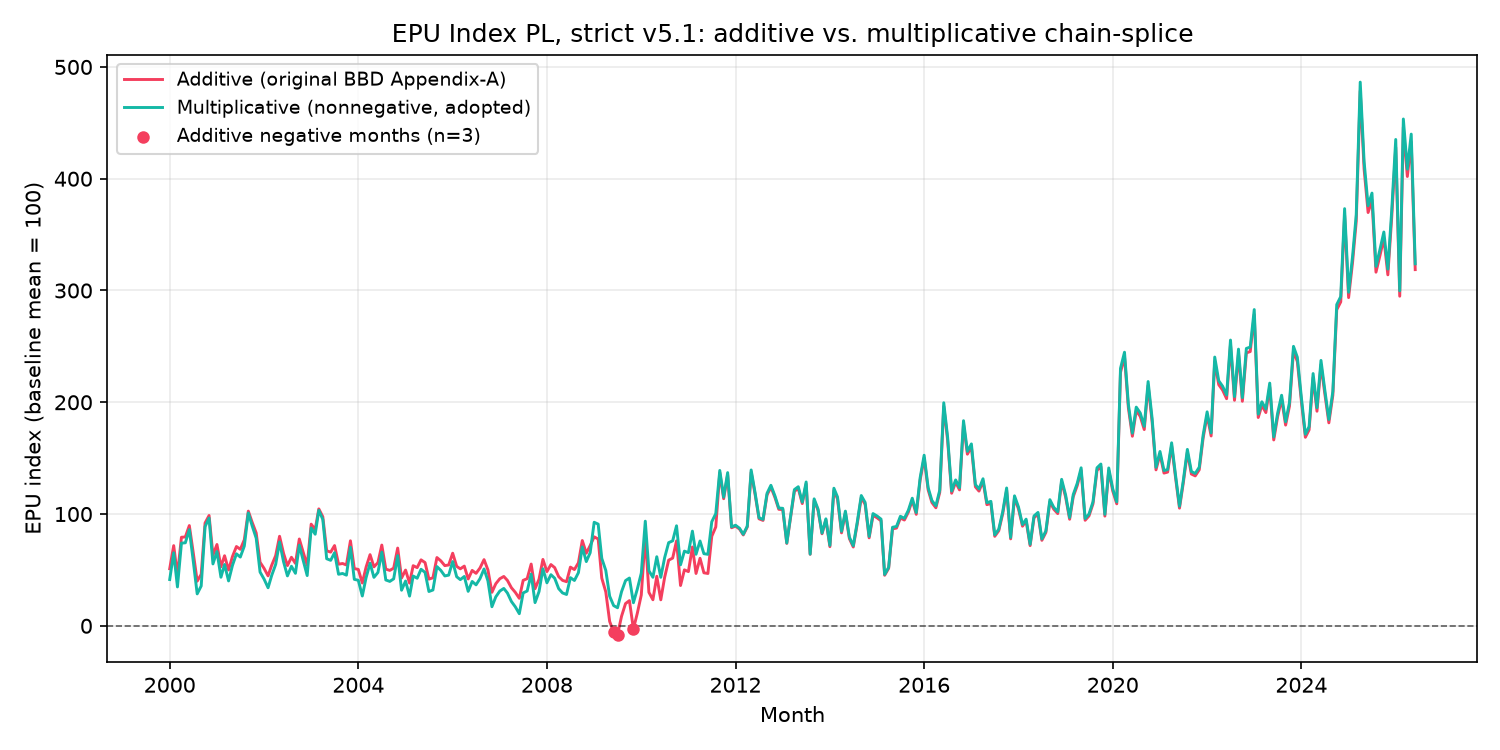}
\caption{EPU Index PL, strict v5.1: additive vs.\ multiplicative chain-splice convention. Markers indicate the three months in which the additive series falls below zero.}
\label{fig:additive_vs_multiplicative}
\wpsource{Authors' calculations.}
\end{figure}

Figure~\ref{fig:additive_vs_multiplicative} plots the v5.1 series under the additive and the multiplicative splice conventions.

Because the multiplicative convention resolves both the negative-value problem and the GFC-attenuation problem identified in Section~\ref{sec:splice_robustness} above, we adopt it as the convention used throughout this paper's main results, and report the additive convention (identical to the original BBD Appendix-A specification) as a labeled robustness variant. 

\begin{figure}[H]
\centering
\includegraphics[width=0.95\textwidth]{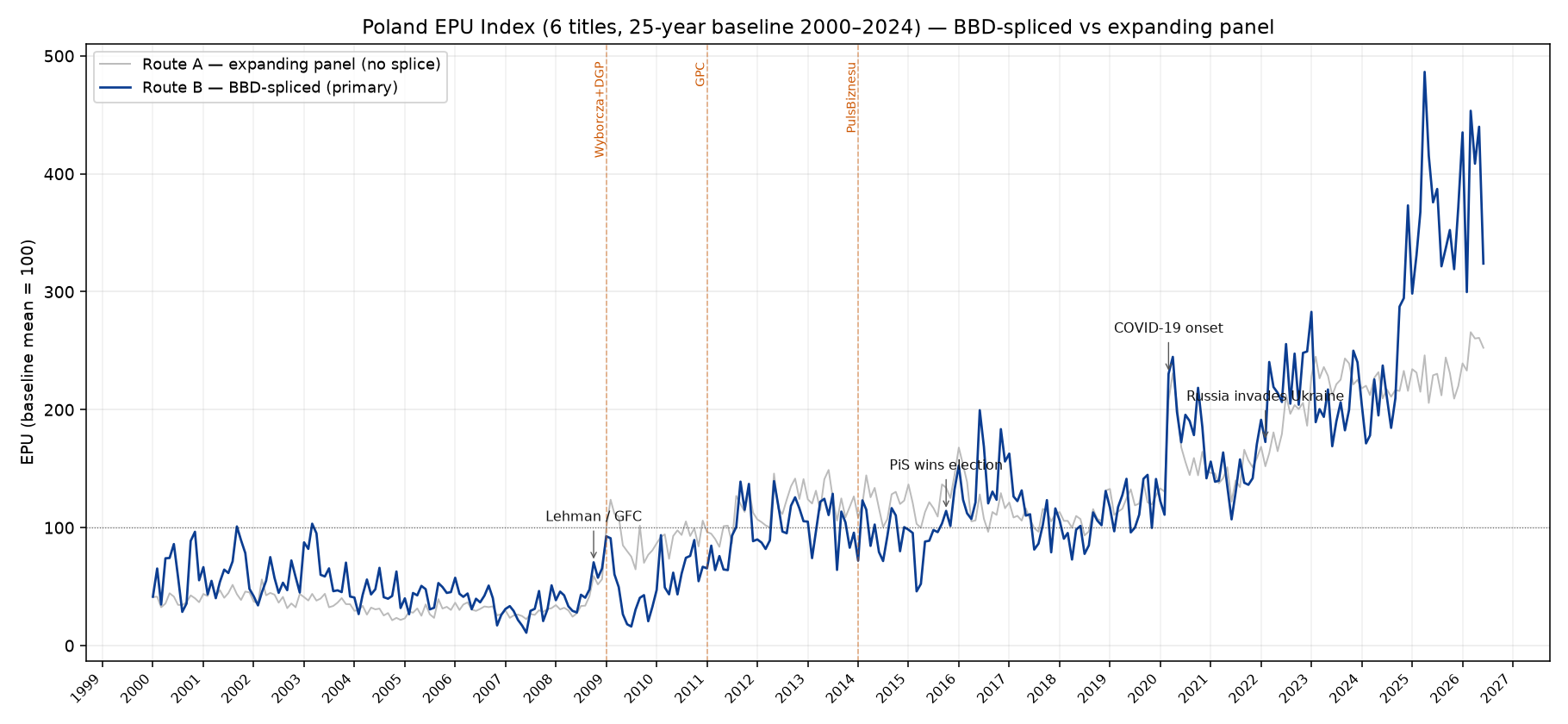}
\caption{Spliced (Route~B) vs.\ expanding-panel (Route~A) monthly EPU Index PL}
\label{fig:spliced}
\wpsource{Authors' calculations.}
\end{figure}

The spliced (Route~B) and expanding (Route~A) monthly series correlate at 0.875 -- high co-movement, with splicing removing the level artifacts at the
three entry boundaries (see Figure~\ref{fig:spliced}).

Because the newspaper panel expands at three points in the sample, comparisons
that are sensitive to panel composition are additionally reported on the
constant-composition window beginning in 2014, when the six-title panel is
complete (see Table~\ref{tab:slant_direct}).

\section{Validation}
\label{sec:validation}

Consistent with the validation approach of the BBD, we assess the performance of the EPU Index PL using four complementary validation exercises.

\subsection{Human audit and precision}
\label{sec:human_audit}

Following BBD, we validate the article-level classification rule through a manual audit conducted by trained annotators. The audit universe comprises all articles satisfying the Economy$\wedge$Uncertainty (E$\wedge$U) criteria, irrespective of whether they also satisfy the Policy criterion; it therefore includes both
articles classified as EPU (E$\wedge$P$\wedge$U) and the near-miss group (E$\wedge$U$\wedge\neg$P). Annotators assess whether each article represents genuine economic policy uncertainty independently of the automated
classification. Comparing these judgments with the algorithmic labels yields the confusion matrix and allows estimation of precision and the gross error rate. The inclusion of near-miss articles enables false negatives of the Policy filter to be identified within the same audit.

\subsubsection{Sample design}
\label{sec:audit_sample_design}

The audit follows the two-stage stratified sampling design of BBD. The sample is drawn algorithmically from the known metadata and classification flags, so the
procedure is exactly reproducible from a fixed seed. Within each newspaper, 20 calendar months are drawn at random from the newspaper's coverage window within
the index period (2000--2026); within each selected month, five articles are sampled without replacement -- four from the flagged E$\wedge$P$\wedge$U cell and
one from the near-miss E$\wedge$U$\wedge\neg$P cell. Sampling by month rather than by year follows the BBD 1985--2012 protocol and insulates the sample from
within-year changes in newspaper size and in the level of economic uncertainty.

The stratification is invisible to annotators: reviewers receive only the article text and neutral bibliographic information, are not told whether an article was
classified as EPU or drawn from the near-miss group, and see the two cells interleaved in randomized order.

Each newspaper contributes 100 audited articles -- 80 flagged and 20 near-miss -- yielding a total sample of 600 articles (480 flagged and 120 near-miss). Because each newspaper series is standardized separately before
aggregation, equal allocation across titles is the natural design-based choice; to characterize the index as actually composed, we additionally report a volume-weighted precision that weights titles by their article counts. The resulting human annotations are then compared with the algorithmic classifications
to estimate precision and the gross error rate.

\subsubsection{Annotators, blinding and adjudication}
\label{sec:adjudication}

Three annotators coded the 600-article audit sample, with 200 articles assigned to each as primary coder. An additional overlap subsample of 120 articles
(20\% of the sample; 40 overlap articles per annotator) was independently double-coded to assess inter-annotator agreement. The total workload was
240 articles per annotator, corresponding to 720 coding decisions.

Annotators were trained using the annotation manual
(see Appendix~\ref{app:annotation_guidelines}) and coded articles independently, without access to the automated classification or matched dictionary terms. For disagreements in the overlap subsample, a third annotator served as adjudicator, and the adjudicated value was used as the final classification. Inter-annotator agreement and adjudication procedure are described in Section~\ref{sec:inter_annotator_agreement}.

\subsubsection{Precision}

Coding proceeds in two steps, with economic policy uncertainty treated as a subset of economic uncertainty: an article is first coded EU, and only articles coded EU$=$1 are eligible for EPU$=$1. The full coding scheme, policy categories and decision rules are set out in Appendix~\ref{app:annotation_guidelines}.

Let EPUC denote the automated flag (EPUC$=$1 on the flagged cell E$\wedge$P$\wedge$U, EPUC$=$0 on the near-miss cell E$\wedge$U$\wedge\neg$P) and EPUH the human classification after adjudication. Cross-tabulating EPUH against EPUC over the audit sample gives the confusion matrix in Table~\ref{tab:confusion_matrix}, with cell counts $\mathit{TP}$, $\mathit{FP}$, $\mathit{FN}$, $\mathit{TN}$.

Because the two cells are sampled at different rates relative to their population sizes, the off-diagonal comparisons require design weights. Let $N^{+}$ and $N^{-}$ be the corpus counts of flagged and near-miss articles under EPU Index PL dictionary, and $n^{+}=480$, $n^{-}=120$ the sample sizes; within the 2000-01--2026-06 index window, over the six audited panel titles only, $N^{+}=23{,}357$ and $N^{-}=4{,}054$, so the weights are $w^{+}=N^{+}/n^{+}=48.66$ and $w^{-}=N^{-}/n^{-}=33.78$, and the weighted counts are $\widetilde{TP}=w^{+}TP$, $\widetilde{FP}=w^{+}FP$, $\widetilde{FN}=w^{-}FN$, $\widetilde{TN}=w^{-}TN$.

Precision is measured on the flagged cell alone and needs no cross-cell weight, since sampling within that cell is uniform:
\begin{equation}
\mathrm{Precision}=\frac{TP}{TP+FP}.
\label{eq:precision}
\end{equation}
Precision is the complement of the false discovery rate ($\mathrm{FDR}=1-\mathrm{Precision}$); it is \emph{not} the false positive rate, which is defined below on a different base. The error rates use both cells with the design weights:
\begin{equation}
\mathrm{FP\ rate}=\frac{\widetilde{FP}}{\widetilde{TP}+\widetilde{FP}+\widetilde{FN}+\widetilde{TN}},\qquad
\mathrm{FN\ rate}=\frac{\widetilde{FN}}{\widetilde{TP}+\widetilde{FP}+\widetilde{FN}+\widetilde{TN}}.
\label{eq:error_rates}
\end{equation}
The false positive and false negative rates share the same (weighted E$\wedge$U) base, so that the \emph{gross error rate}, following BBD, is their sum:
\begin{equation}
\mathrm{Gross\ error}=\mathrm{FP\ rate}+\mathrm{FN\ rate}.
\label{eq:gross_error}
\end{equation}
This is the criterion BBD minimize when selecting the Policy term set; here it provides the diagnostic for comparing dictionary versions (see Table~\ref{tab:gross_error_versions}).

Confidence intervals for the design-based (equal-allocation) precision use the Wilson score interval, which is better behaved than the normal approximation near the boundary and for smaller per-newspaper cells. The volume-weighted precision's confidence interval instead uses a normal approximation around the weighted point estimate, with variance given by the weighted sum of each title's binomial variance (design-based standard error): Wilson's score interval does not generalize to a weighted combination of six independent binomial proportions, so the two precision rows in Table~\ref{tab:human_audit_results} use different interval methods. Tables~\ref{tab:confusion_matrix} and~\ref{tab:human_audit_results} report the outcomes after adjudication.

\begin{table}[H]
\centering
\caption{Human vs.\ automated classification over the E$\wedge$U universe}
\label{tab:confusion_matrix}
\small
\begin{tabular}{lccc}
\toprule
 & \textbf{EPUH$=$1} & \textbf{EPUH$=$0} & \textbf{Total} \\
\midrule
\textbf{EPUC$=$1 (flagged, E$\wedge$P$\wedge$U)}   & $TP=337$ & $FP=143$ & 480 \\
\textbf{EPUC$=$0 (near-miss, E$\wedge$U$\wedge\neg$P)}  & $FN=34$ & $TN=86$ & 120 \\
\midrule
\textbf{Total}                                     & 371 & 229 & 600 \\

\bottomrule
\end{tabular}
\vspace{3pt}
\begin{flushleft}
\footnotesize
\textit{Notes:} Classification after adjudication. EPUC is the automated flag; EPUH the human coding. Row totals are the sampled cell sizes (480 flagged, 120 near-miss); the error rates apply the population weights $w^{+},w^{-}$ defined in the text.
\end{flushleft}
\end{table}

\begin{table}[H]
\centering
\caption{Human audit results}
\label{tab:human_audit_results}
\small
\begin{tabular}{lrr}
\toprule
\textbf{Quantity} & \textbf{Estimate} & \textbf{95\% CI} \\
\midrule
Precision (flagged cell, design-based)   & 70.2\% & 66.0--74.1\% \\
Precision (volume-weighted)              & 70.6\% & 65.9--75.2\% \\
FP rate (weighted E$\wedge$U base)       & 25.4\% & --- \\
FN rate (weighted E$\wedge$U base)       & 4.2\% & --- \\
Gross error rate (FP$+$FN)               & 29.6\% & --- \\
\bottomrule
\end{tabular}
\vspace{3pt}
\begin{flushleft}
\footnotesize
\vspace{3pt}
\textit{Notes:} Precision on the flagged cell is reported both
design-based (70.2\%, Wilson 95\% CI 66.0--74.1\%, $n_{+}=480$)
and volume-weighted (70.6\%, weighting the 6 audited titles by their
share of flagged articles in the full corpus; see
Section~\ref{sec:adjudication}). Precision restricted to articles with coding
confidence $\geq 2$ is 69.1\% ($n=466$). FP and FN rates share the
weighted $E\wedge U$ base, so the gross error rate is their sum
(see Equations~\eqref{eq:precision}--\eqref{eq:gross_error}). Breakdowns by
newspaper and by period are reported in Appendix~\ref{app:audit_breakdown}.
\end{flushleft}
\end{table}

Table~\ref{tab:gross_error_versions} reports the gross error rate of each successive dictionary version, evaluated against the adjudicated human labels on the same 600-article audit sample. Gross error falls from 36.6\% for v1 to 29.6\% for v5, which attains the minimum among the versions compared: each iteration trades a moderate rise in the false-positive rate for a larger fall in the false-negative rate (from 16.5\% to 4.2\%). The comparison thus corroborates the iterative refinement described in Section~\ref{sec:epu_framework}.

\begin{table}[H]
\centering
\caption{Gross error rate by dictionary version (same 600-article audit sample)}
\label{tab:gross_error_versions}
\small
\begin{tabular}{lrrrrrrr}
\toprule
\textbf{Version} & \textbf{TP} & \textbf{FP} & \textbf{FN} & \textbf{TN} & \textbf{FP rate (\%)} & \textbf{FN rate (\%)} & \textbf{Gross error (\%)} \\
\midrule
v1   & 263 & 111 & 105 & 121 & 20.1 & 16.5 & 36.6 \\
v2   & 297 & 120 &  71 & 112 & 21.7 & 10.3 & 32.0 \\
v3   & 311 & 135 &  57 &  97 & 22.4 &  8.9 & 31.3 \\
v4.1 & 306 & 124 &  62 & 108 & 22.5 &  8.7 & 31.2 \\
v4.2 & 306 & 127 &  62 & 105 & 23.0 &  8.7 & 31.7 \\
v5   & 337 & 143 &  34 &  86 & 25.4 &  4.2 & 29.6 \\
\bottomrule
\end{tabular}
\begin{flushleft}
\footnotesize
\textit{Notes:} Each version's classification of the 600 audit articles is compared with the adjudicated human label (EPUH). FP and FN rates are design-weighted, as in Table~\ref{tab:human_audit_results}; gross error is their sum. Among the six versions evaluated against this audit sample, v5 attains the lowest gross error and was adopted as the base dictionary; the published index additionally uses v5.1 (v5 plus four European-institution terms, Table~\ref{tab:dictionary_diff}), which does not require a separate row here since it classifies the audit sample identically to v5 (Section~\ref{sec:epu_framework}).
\end{flushleft}
\wpsource{Authors' calculations.}
\end{table}

The taxonomy of false positives in Table~\ref{tab:false_positive_taxonomy}
identifies the main sources of filter errors and informs future dictionary refinement.

\begin{table}[H]
\centering
\caption{Composition of false positives}
\label{tab:false_positive_taxonomy}
\small
\begin{tabular}{p{8.4cm}rr}
\toprule
\textbf{Category of error} & \textbf{Articles} & \textbf{Share of FP (\%)} \\
\midrule
(1) Uncertainty unrelated to economic policy                    & 60 & 42.0 \\
(2) Policy terms referring to non-economic domains              & 57 & 39.9 \\
(3) Economic terms used in a non-policy context                 & 12 & 8.4 \\
(4) Institutional terms referring to actors or events unrelated
    to economic policy decisions                                & 3 & 2.1\\
(5) E, P and U elements present but not connected               & 11 & 7.7 \\
\midrule
Total false positives                                           & 143 & 100.0 \\

\bottomrule
\end{tabular}

\vspace{3pt}
\begin{flushleft}
\footnotesize
\textit{Notes:} Error categories follow the taxonomy in Appendix~\ref{app:fp_taxonomy}. Categories are assigned by the annotator whose classification was retained after adjudication. Following the 2026-10-20 redraw of 20 out-of-window Rzeczpospolita articles (Section~\ref{sec:adjudication}), 7 of the 20 replaced articles had been false positives (three in category 1, two in category 2, and one each in categories 4 and 5); 3 of the 20 replacement articles are themselves false positives (one each in categories 1, 3 and 5, assigned by adjudication). Total FP for v5 therefore fell from 147 to 143. The distribution identifies candidate revisions to the dictionaries and complements the gross-error comparison across dictionary versions carried out within the same audit (see Table~\ref{tab:gross_error_versions}).
\end{flushleft}
\end{table}

\subsubsection{Inter-annotator agreement}
\label{sec:inter_annotator_agreement}

Agreement is assessed on the overlap subsample of 120 articles. Because Cohen's $\kappa$ is sensitive to the prevalence of the positive class, we report observed agreement alongside Cohen's $\kappa$ and the prevalence-adjusted measures PABAK and Gwet's $AC_1$. The resulting agreement statistics are presented in Table~\ref{tab:agreement_stats}, while the full agreement matrices are reported in Appendix~\ref{app:inter_annotator}.

\begin{table}[!htbp]
\centering
\caption{Inter-annotator agreement statistics}
\label{tab:agreement_stats}
\small
\begin{tabular}{lccccc}
\toprule
\textbf{Pair} & \textbf{$n$} & \textbf{Observed (\%)} &
\textbf{Cohen's $\kappa$} & \textbf{95\% CI} & \textbf{$AC_1$} \\
\midrule
A1--A2 & 40  & 75.0 & 0.47 & 0.16--0.73 & 0.53 \\
A1--A3 & 40  & 92.5 & 0.83 & 0.62--1.00 & 0.86 \\
A2--A3 & 40  & 95.0 & 0.90 & 0.74--1.00 & 0.90 \\
\midrule
Pooled & 120 & 87.5 & 0.74 & 0.60--0.85 & 0.76 \\
\bottomrule
\end{tabular}

\vspace{3pt}
\begin{flushleft}
\footnotesize
\textit{Notes:} Agreement on the binary EPU field, before adjudication. Pooled Fleiss' $\kappa=0.74$; PABAK $=0.75$. The prevalence of valid EPU cases in the
overlap subsample is 61.25\%; at this prevalence $\kappa$ is attenuated relative to observed agreement, which is why prevalence-adjusted statistics are reported
alongside. Agreement restricted to articles for which both coders assigned coding confidence $\geq 2$ is 89.7\% ($n=$~107 pairwise comparisons).
\end{flushleft}
\end{table}

\subsubsection{Limitations}

\label{sec:audit_limitations}

Three limitations qualify these results. First, the near-miss cell supplies the false negatives that enter the gross error rate and the dictionary-version comparison, but with 120 near-miss articles it is not sized to yield a population-level recall estimate with a usable confidence interval; we therefore report the gross error rate rather than recall, and leave a precise recall
estimate to a larger audit. For the same reason, the false-negative rate -- which weights the near-miss cell up to its population size and so depends on the corpus
counts $N^{+},N^{-}$ -- is more sensitive to sampling variability than precision, which is measured directly on the flagged cell.

Second, precision is estimated on the 480 flagged articles with a 95\% confidence interval of roughly $\pm$3--4 percentage points; estimates by newspaper ($n=80$ flagged per title) and by period ($n=64$--$232$) rest on smaller cells, with intervals of roughly $\pm$6 to $\pm$11 percentage points, and should be read as descriptive.

Third, the gross error rate lets us compare dictionary versions on the audit sample (see Table~\ref{tab:gross_error_versions}), but a dictionary selected or revised on this sample cannot then be evaluated on it without overstating its performance; any revision motivated by the false-positive taxonomy or the near-miss analysis must be validated on an independent sample.

\subsection{Historical validation: Known uncertainty episodes}
\label{sec:events}

\begin{figure}[H]
\centering
\includegraphics[width=0.97\textwidth]{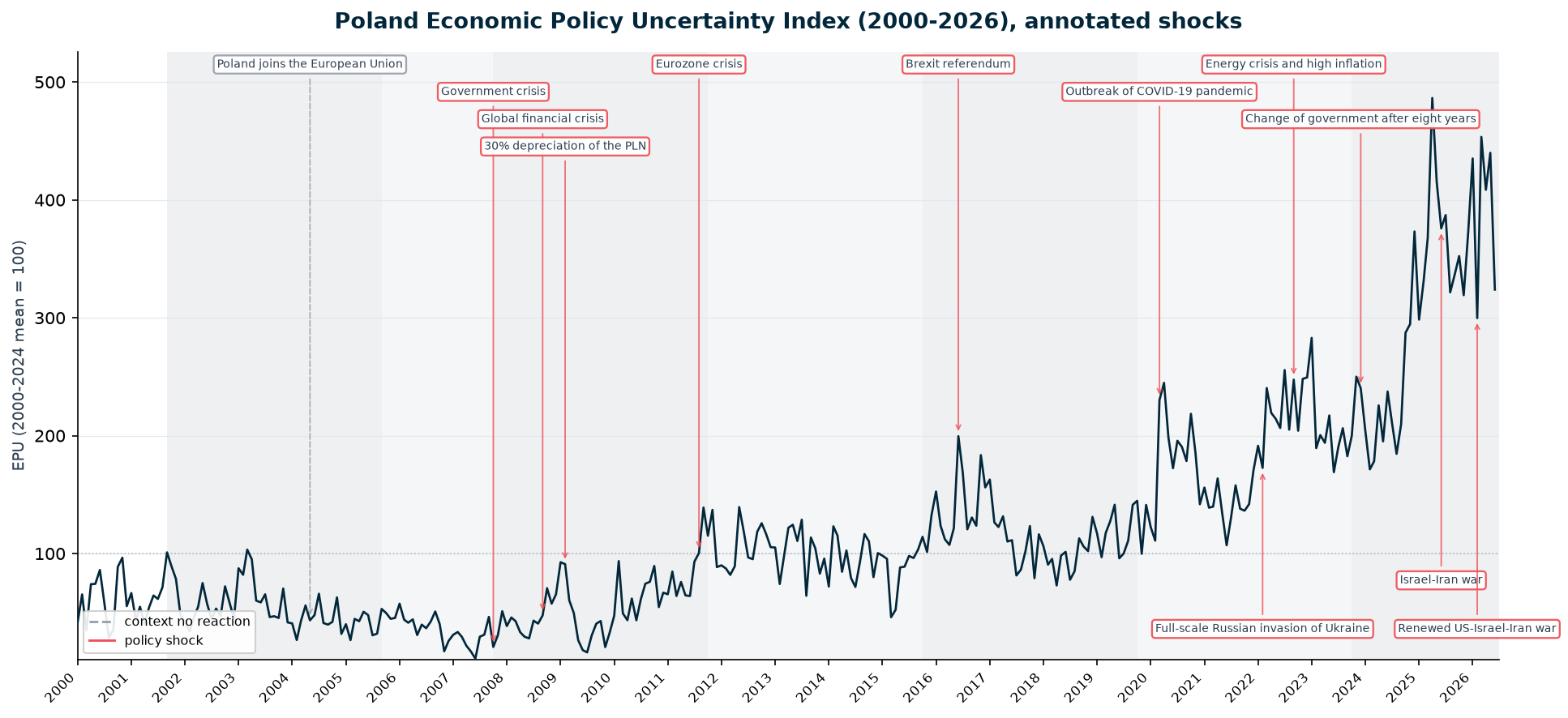}
\caption{Annotated EPU Index PL narrative: Reacting shocks (solid) and important-but-no-reaction events (dashed)}
\label{fig:narrative}
\wpsource{Authors' calculations.}
\end{figure}

Following BBD, the index is evaluated on two criteria. \emph{Face validity} requires its largest movements to coincide with independently recognized episodes of elevated economic policy uncertainty. \emph{Discriminant validity} requires that major news events that lack an economic-policy component do \emph{not} trigger comparable movements, distinguishing the index from a measure of media attention. Because the series exhibits an upward trend and peak magnitudes depend on panel composition (Section~\ref{sec:splice_robustness}), episodes are evaluated by their deviation from a local baseline ($z$), with raw levels reported for reference. A reaction is recorded when $z \geq 1$. Figure~\ref{fig:narrative} shows the annotated series.

\subsubsection{Discriminant validity: non-policy controls}
\label{sec:events_controls}

Three heavily covered events without a discrete economic-policy component serve as controls. The Smolensk air disaster (2010-04), which killed the President, the governor of the central bank and much of the senior state leadership, leaves the index near its local baseline ($z=-0.5$). The judicial reform and rule-of-law dispute (2017-07)---a prolonged constitutional conflict dominating domestic and European news coverage---remains below baseline ($z=-1.6$). Poland's accession to the European Union (2004-05), although a landmark event, produces virtually no reaction ($z=+0.1$).

This provides a stringent discriminant-validity test: despite extensive media attention, none of these events raises the index. In particular, the judicial reform episode shows that political conflict alone is insufficient; the index responds only when an economic-policy uncertainty channel is present.

\subsubsection{Face validity: domestic versus external sources of uncertainty}
\label{sec:events_face}

The clearest reactions occur during episodes of \emph{domestic} economic-policy uncertainty (see Table~\ref{tab:events}): the COVID-19 lockdown (2020-03, $z=+1.8$), the change of government after eight years (2023-12, $z=+1.7$), and the energy-price and inflation shock (2022-09, $z=+1.1$). In each case, uncertainty concerns fiscal or monetary policy, the authority responsible for economic policy, or the regulatory framework. A positive local deviation is also observed during the sharp depreciation of the zloty in February 2009 ($z=+1.0$), an episode of financial-market stress that may have heightened uncertainty about domestic economic-policy responses. The Brexit referendum (2016-06) produces the largest local deviation among the selected episodes ($z=+3.2$). Although the referendum took place outside Poland, its potential implications for Polish trade, migration and EU funding illustrate why EU-level policy developments are relevant to the Polish economic-policy environment and motivate the inclusion of EU-related institutional terms in the Policy dictionary.

By contrast, externally driven shocks exhibit a different pattern. The Global Financial Crisis (2008-09, $z=+0.3$), Russia's full-scale invasion of Ukraine (2022-02, $z=-0.7$), and the 2025 Middle East escalation (2025-06, $z=+0.3$) do not generate pronounced positive local deviations. This contrasts with the sharper responses observed during selected domestic economic-policy episodes. The distinction concerns deviations from the prevailing local baseline rather than the absolute level of the index.

This asymmetry is consistent with EPU Index PL primarily capturing uncertainty about domestic economic policy rather than geopolitical risk or international shocks as such. Under coding rule~R8 (Appendix~\ref{app:annotation_guidelines}), external events enter the index when newspaper coverage connects them to uncertainty about economic-policy decisions affecting Poland. Their effects may therefore be reflected in domestic policy responses rather than in the immediate month of the external shock.

\begin{table}[H]
\centering
\caption{Known-event validation. Level: index value in the episode month;
$z$: deviation from local baseline; a reaction is recorded when $z \geq +1$}
\label{tab:events}
\small
\begin{tabular}{lllccc}
\toprule
Month & Event & Type & Level & $z$ & Reaction \\
\midrule
\multicolumn{6}{l}{\emph{Economic-policy uncertainty episodes affecting Poland}} \\
2016-06 & Brexit referendum                & policy & 196.9 & $+3.2$ & yes \\
2020-03 & COVID-19 lockdown                & policy & 227.1 & $+1.8$ & yes \\
2023-12 & Change of government             & policy & 236.9 & $+1.7$ & yes \\
2022-09 & Energy crisis and high inflation & policy & 244.1 & $+1.1$ & yes \\
2009-02 & 30\% depreciation of the PLN     & policy & 77.8  & $+1.0$ & yes \\
2015-10 & PiS wins parliamentary majority  & policy & 112.6 & $+0.5$ & no \\
2011-08 & Eurozone crisis                  & policy & 88.5  & $+0.2$ & no \\
\midrule
\multicolumn{6}{l}{\emph{External / geopolitical shocks}} \\
2025-06 & Israel--Iran war                 & external & 375.8 & $+0.3$ & no \\
2008-09 & Lehman / Global Financial Crisis & external & 47.5  & $+0.3$ & no \\
2022-02 & Russia invades Ukraine           & external & 166.7 & $-0.7$ & no \\
2026-02 & Renewed US--Israel--Iran war     & external & 295.4 & $-1.2$ & no \\
\midrule
\multicolumn{6}{l}{\emph{Non-policy controls}} \\
2004-05 & Poland joins the EU               & control & 52.8 & $+0.1$ & no \\
2010-04 & Smolensk air disaster             & control & 23.4 & $-0.5$ & no \\
2017-07 & ``Judicial reform'' / rule-of-law & control & 80.3 & $-1.6$ & no \\
\bottomrule
\end{tabular}
\begin{flushleft}
\footnotesize
\textit{Notes:} $z$ is the standardized deviation of the episode month from a local baseline; a reaction requires $z \geq +1$, so declines below baseline (2026-02, 2017-07) are not recorded as reactions. The ``Level'' column is relative to the fixed 2000--2024 baseline mean of 100 and is not directly comparable across early- and late-sample episodes, since the index's absolute level rises over the sample; the 2008-09 GFC level (47.5) is below the fixed baseline mean of 100 even though it is elevated ($z=+0.3$) relative to its own local (pre-GFC) trend; the 2025-06 episode shown here is a large but not record-setting reaction -- the series' true sample maximum of 486.4 occurs in 2025-04 (see Table~\ref{tab:additive_vs_multiplicative}) -- the abstract's and Section~\ref{sec:events}'s claim that the GFC ``raises its level'' refers to this local-trend sense, not to the fixed baseline. Readings for events embedded in an already-elevated regime are sensitive to the definition of the baseline window. \textit{Source:} Authors' calculations.
\end{flushleft}
\end{table}

\FloatBarrier
\subsection{Media slant and partisan robustness}
\label{sec:slant}

Newspaper measures can be distorted if coverage of uncertainty tracks a title's editorial orientation rather than the phenomenon: a title may amplify uncertainty when a government it opposes is in office.
\citet{baker2016measuring} test for this by comparing indices from ideologically opposed titles and by checking whether a title's relative coverage moves with the party in power. The Polish panel contains one
consistently liberal title (\emph{Gazeta Wyborcza}, GW) and one consistently conservative title (\emph{Gazeta Polska Codziennie}, GPC); the other four are business or quality titles without partisan alignment (see Appendix~\ref{app:slant}, Table~\ref{tab:slant_classification}). Each newspaper-level series uses the same E$\wedge$P$\wedge$U rule and
normalization as the aggregate. We report three tests.

\subsubsection{Direct comparison of ideologically opposed titles}
\label{sec:slant_direct}

Table~\ref{tab:slant_direct} correlates the left-of-center (GW) and right-of-center (GPC) series directly. In levels the two titles correlate at $r=0.66$, close to the $0.67$ that \citet{holda2019newspaper} reports for GW against \emph{Rzeczpospolita} and consistent with both titles tracking the same underlying episodes. In month-on-month changes the correlation is small ($0.07$; $0.10$ on quarterly data, which re-aggregates counts before differencing). This is not specific to the left-right pair: in the full title-by-title matrix (see Appendix~\ref{app:slant}, Table~\ref{tab:slant_correlation}) \emph{every} pair of single titles correlates weakly in changes; the strongest change-correlations are all among the business and economic titles -- the highest off-diagonal change-correlation is between two business titles (\emph{Parkiet}$\times$\emph{Puls Biznesu}, $0.34$). The weak change-correlation therefore reflects the idiosyncratic sampling noise of any single title's monthly $C/N$ ratio, not editorial divergence; whether the divergence is \emph{political} is settled by the amplification test below, not by the size of this correlation. The common window runs 2011-09--2026-06 (GPC enters 2011-09, 178 months inclusive at monthly frequency, 177 after first-differencing), the smallest number of months common to both series; the 2014+ window is the constant-composition check.

\begin{table}[!htbp]
\centering
\caption{Direct correlation of ideologically opposed titles}
\label{tab:slant_direct}
\small
\begin{tabular}{llcccc}
\toprule
\textbf{Comparison} & \textbf{Window} & \textbf{Transform} &
\textbf{Pearson} & \textbf{Spearman} & \textbf{$n$} \\
\midrule
GW $\times$ GPC & common & levels             & 0.664 & 0.569 & 178 \\
\addlinespace
GW $\times$ GPC & common & changes, monthly    & 0.072 & 0.128 & 177 \\
GW $\times$ GPC & 2014+  & changes, monthly    & 0.095 & 0.161 & 150 \\
GW $\times$ GPC & common & changes, quarterly  & 0.103 & 0.169 & 59 \\
GW $\times$ GPC & 2014+  & changes, quarterly  & 0.149 & 0.218 & 50 \\
\addlinespace
GW $\times$ core  & common & changes, monthly   & 0.058 & 0.089 & 178 \\
GPC $\times$ core & common & changes, monthly   & 0.100 & 0.058 & 177 \\
\bottomrule
\end{tabular}
\begin{flushleft}
\footnotesize
\textit{Notes:} Non-partisan core = volume-weighted series of
\emph{Rzeczpospolita}, \emph{Parkiet}, \emph{Dziennik Gazeta Prawna} and
\emph{Puls Biznesu}. Changes are first differences; the quarterly rows
re-aggregate the raw article counts by quarter before forming the ratio.
$n$ is the number of jointly available observations. For pairs not involving GPC, the first monthly change in the common window is computed from the pre-window month, so $n$ equals the number of months in the window.
\end{flushleft}
\end{table}

\subsubsection{Partisan amplification test}
\label{sec:slant_partisan}

For each partisan title $p$ we form its deviation from the aggregate computed \emph{excluding that title} (leave-one-out, removing any mechanical dependence),
\begin{equation}
d_{p,t}=z_{p,t}-z_{\text{agg}\setminus p,\,t},
\label{eq:slant_deviation}
\end{equation}
with $z$ standardized over the estimation window. Over 2011--2026 the governing coalition changes twice: PO--PSL to PiS in November~2015, and PiS to a KO-led coalition in December~2023. With PO--PSL as the omitted
category,
\begin{equation}
d_{p,t}=\alpha+\beta_{1}\,\mathrm{PiS}_{t}+\beta_{2}\,\mathrm{KO}_{t}
+\varepsilon_{p,t},
\label{eq:slant_regression}
\end{equation}
estimated separately for GW and GPC with Newey--West standard errors. Slant implies $\beta_{1}>0$ for GW and $\beta_{1}<0$ for GPC. Column~(3) pools the two titles on an indicator $\mathrm{Opp}_{p,t}$ equal to one when the title's side is in opposition; under no slant its coefficient is zero.

\begin{table}[!htbp]
\centering
\caption{Partisan amplification: deviation of partisan titles from the
leave-one-out aggregate}
\label{tab:slant_partisan}
\small
\begin{tabular}{lccc}
\toprule
 & \textbf{(1) GW} & \textbf{(2) GPC} & \textbf{(3) Pooled} \\
\midrule
PiS government ($\beta_1$)           & $-0.433^{**}$ & $-0.256^{*}$ & --- \\
KO government ($\beta_2$)            & $-0.198$ & $+0.104$ & --- \\
Side in opposition ($\mathrm{Opp}$)  & --- & --- & $0.003$ \\
\addlinespace
Constant                             & $0.211$ & $0.187$ & $0.194$ \\
Observations                         & 165 & 165 & 330 \\
$R^2$                                & 0.041 & 0.028 & 0.019 \\

\bottomrule
\end{tabular}
\begin{flushleft}
\footnotesize
\textit{Notes:} Dependent variable is the standardized deviation of the
title from the aggregate excluding that title. Omitted category: PO--PSL.
Significance is based on Newey--West standard errors: $^{**}$~$p<0.05$; $^{*}$~$p<0.10$. Coefficients indistinguishable
from zero indicate no partisan amplification. The slant hypothesis predicts $\beta_1>0$ for GW and $\beta_1<0$ for GPC (Section~\ref{sec:slant}); GW's coefficient is significant but of the opposite sign to this prediction and therefore does not support amplification by GW, while GPC's is significant and correctly signed.
\end{flushleft}
\end{table}

\subsubsection{Robustness to the exclusion of partisan titles}
\label{sec:slant_leaveout}

Table~\ref{tab:slant_leaveout} recomputes the aggregate excluding both partisan titles and correlates it with the published index; it also reports the leave-one-out variants across all six titles. This check covers the
pre-2011 period the GW--GPC comparison cannot, since the non-partisan titles reach back to 2000/2009.

\begin{table}[!htbp]
\centering
\caption{Sensitivity of the aggregate index to newspaper composition}
\label{tab:slant_leaveout}
\small
\begin{tabular}{lcc}
\toprule
\textbf{Aggregate variant} & \textbf{Corr.\ with published (changes)}
& \textbf{Max.\ level gap} \\
\midrule
Excluding GW and GPC (non-partisan core) & 0.800 & 87.7 \\
Excluding GW only                        & 0.830 & 93.6 \\
Excluding GPC only                       & 0.898 & 33.9 \\
Minimum across all leave-one-out variants & 0.784 & 140.1 \\
\bottomrule
\end{tabular}
\begin{flushleft}
\footnotesize
\textit{Notes:} Each variant re-standardizes and re-averages the retained titles following the baseline procedure, over the full common sample (2000-01 to 2026-06, $n=317$ after first-differencing the 318-month level series to month-on-month changes). The final column is the largest absolute difference in level (baseline~$=100$) from the published series. The minimum across all six single-title leave-one-out variants is Rzeczpospolita (corr.\ 0.784, max.\ gap 140.1), the largest drop of any single-title exclusion -- expected, since Rzeczpospolita is one of only two titles (with Parkiet) spanning the full 2000--2008 pre-entry period.
\end{flushleft}
\end{table}

\noindent
Ideologically opposed titles co-move at $r=0.68$ in levels and $r=0.07$ in changes (see Table~\ref{tab:slant_direct}); relative coverage does not shift with the governing coalition in the pooled specification ($\mathrm{Opp}$ coefficient indistinguishable from zero,
Table~\ref{tab:slant_partisan}); and removing either partisan pole leaves the index materially unchanged (correlation~$\geq0.80$, maximum level gap~$\leq94$ points,
Table~\ref{tab:slant_leaveout}). The per-title amplification coefficients in Table~\ref{tab:slant_partisan} are individually significant for both GW and (at 10\%) GPC, but only GPC's coefficient has the sign the slant hypothesis predicts ($\beta_1<0$); GW's coefficient ($-0.433^{**}$) is significant but \emph{sign-opposite} to the hypothesis ($\beta_1>0$ was predicted), so it does not constitute evidence of amplification by GW in the direction the slant story requires. The pooled opposition-indicator test -- the cleaner test of a partisan-amplification story common to both titles -- shows no significant effect. Taken together, we read this as, at most, weak and title-specific evidence of slant restricted to GPC, with no evidence of amplification by GW and no systematic partisan distortion of the aggregate index. The full title-by-title correlation matrix is in Appendix~\ref{app:slant}, Table~\ref{tab:slant_correlation}.

\subsection{Validation Against Other Measures of Uncertainty}
\label{sec:correlations}

To assess the external validity and economic relevance of EPU Index PL, we examine its relationship with established measures of policy uncertainty, geopolitical risk and financial-market conditions. The comparison is organized around four dimensions: (i) global policy and geopolitical uncertainty, (ii) financial-market uncertainty, (iii) domestic financial transmission channels, and (iv) safe-haven and alternative risk assets.

The correlations between the EPU Index PL and the benchmark indicators are reported in Table~\ref{tab:epu_correlations}, computed on month-on-month percentage changes of both EPU (strict v5.1, multiplicative-splice vintage; Section~\ref{sec:additive_vs_multiplicative}) and each indicator over the longest window for which each indicator has native monthly data (2000+ unless noted; see table notes). Appendix Table~\ref{tab:all_correlations} additionally reports every lag ($t$/$t$-1/$t$-2/$t$-3) at the indicator's first supported window; the full set of all four evidence-selected windows (2000+, 2008+, 2014+ and 2020+), the levels-transform comparisons, and exact p-values for every cell are provided in the online replication files rather than reproduced in the appendix.

Correlations computed on raw levels of EPU and the benchmark indicators are in several cases unusually strong and difficult to reconcile with expected economic relationships. For example, Gold and Bitcoin exhibit positive level correlations of $r=+0.89$ and $r=+0.83$, respectively, while the WIG correlates positively with EPU at $r=+0.76$. Such correlations may reflect common trends rather than short-run economic co-movement \citep{granger1974spurious}. We therefore use month-on-month percentage changes to reduce the influence of these trends and focus on short-run dynamics. Under this transformation, Gold's correlation falls to $r=+0.030$ ($p=0.592$), while the WIG correlation changes sign to $r=-0.203$. Table~\ref{tab:epu_correlations} reports correlations in monthly changes as the primary results, while levels-based correlations are retained in the replication materials for comparison. Because native monthly coverage differs by indicator, and because the
reported lag itself varies by row ($t$, $t\text{-}1$, $t\text{-}2$ or
$t\text{-}3$), the effective sample size underlying
Table~\ref{tab:epu_correlations} is not constant even within the nominal
2000+ window: it is $n=317$ for indicators with full native
2000-01--2026-06 coverage reported at lag $t$ (Global EPU Index, US EPU
Index, GPR Index, VIX, MOVE, V2X, EUR/PLN, EuroStoxx~50) or at lag
$t\text{-}1$ (WIBOR~6M, WIBOR~3M) -- $n$ does not fall when the reported
lag moves from $t$ to $t\text{-}1$ for these two rows specifically,
because their own native history extends one month beyond EPU's 2026-06
end (2026-07), so a one-month shift still lands within the overlapping
sample; $n$ falls to $n=315$--$316$ for rows whose own native
history ends slightly earlier (USD/PLN, WIG, WIG20, Gold: last observation
2026-04) or that are reported at a longer lag (PL~10Y yield, $t\text{-}2$,
$n=316$), and is smaller still for indicators whose native coverage starts
later than 2000-01 (WIBOR~1Y: $n=222$ at 2008+; Polish CDS: $n=210$ at
2008+; WIRON: $n=76$ at 2020+; Bitcoin: $n=148$ at 2014+). The exact $n$ underlying every cell is provided in the replication files.

\begin{table}[!htbp]
\centering
\caption{Correlation between EPU Index PL and uncertainty, financial and macroeconomic indicators (month-on-month \% changes)}
\label{tab:epu_correlations}
\small

\begin{tabular}{p{0.22\textwidth} p{0.33\textwidth} p{0.12\textwidth} p{0.10\textwidth} p{0.10\textwidth}}
\toprule
\textbf{Indicator} &
\textbf{Description} &
\textbf{Window} &
\textbf{Best lag} &
\textbf{Pearson $r$}
\\
\midrule

\multicolumn{5}{l}{\textbf{Global uncertainty benchmarks}}\\
Global EPU Index
& Global Economic Policy Uncertainty
& 2000+
& t
& +0.248\textsuperscript{***}
\\

US EPU Index
& United States EPU
& 2000+
& t
& +0.258\textsuperscript{***}
\\

\midrule

\multicolumn{5}{l}{\textbf{Geopolitical uncertainty}}\\
GPR Index
& Geopolitical Risk Index (30-day MA)
& 2000+
& t
& +0.066
\\

\midrule

\multicolumn{5}{l}{\textbf{Financial volatility}}\\
VIX
& US equity volatility
& 2000+
& t
& +0.257\textsuperscript{***}
\\

MOVE
& US bond volatility
& 2000+
& t
& +0.196\textsuperscript{***}
\\

V2X
& Euro area equity volatility
& 2000+
& t
& +0.277\textsuperscript{***}
\\

\midrule

\multicolumn{5}{l}{\textbf{Domestic financial conditions}}\\
EUR/PLN
& Polish exchange rate
& 2000+
& t
& +0.142\textsuperscript{**}
\\

USD/PLN
& US dollar exchange rate
& 2000+
& t
& +0.148\textsuperscript{***}
\\

\midrule

\multicolumn{5}{l}{\textbf{Interest rates and monetary conditions}}\\
WIBOR 1Y
& Polish money market rate
& 2008+\textsuperscript{a}
& t-1
& $-$0.059
\\

WIBOR 6M
& Polish money market rate
& 2000+
& t-1
& $-$0.040
\\

WIBOR 3M
& Polish money market rate
& 2000+
& t-1
& $-$0.037
\\

WIRON
& Polish overnight reference rate
& 2020+\textsuperscript{a}
& t-3
& +0.096
\\

PL 10Y government yield
& Sovereign bond market
& 2000+
& t-2
& $-$0.087
\\

\midrule

\multicolumn{5}{l}{\textbf{Sovereign risk}}\\
Polish CDS / sovereign spread
& Country risk premium
& 2008+\textsuperscript{a}
& t
& +0.211\textsuperscript{***}
\\

\midrule

\multicolumn{5}{l}{\textbf{Equity markets}}\\
WIG
& Warsaw broad equity index
& 2000+
& t
& $-$0.203\textsuperscript{***}
\\

WIG20
& Warsaw blue-chip index
& 2000+
& t
& $-$0.164\textsuperscript{***}
\\

EuroStoxx 50
& European equities
& 2000+
& t
& $-$0.233\textsuperscript{***}
\\

\midrule

\multicolumn{5}{l}{\textbf{Safe-haven and risk-sensitive assets}}\\
Gold
& Safe-haven commodity
& 2000+
& t
& +0.030
\\

Bitcoin
& Risk-sensitive speculative asset
& 2014+\textsuperscript{a}
& t-3
& $-$0.071
\\

\bottomrule
\end{tabular}
\begin{flushleft}
\footnotesize
\textit{Notes:} All correlations computed on month-on-month \% changes
(EPU: \texttt{epu\_strict\_v51\_mult}, multiplicative-splice vintage). Lag
``t-$k$'' means EPU leads the indicator by $k$ months. \textsuperscript{***}
$p<0.01$; \textsuperscript{**} $p<0.05$; \textsuperscript{*} $p<0.10$;
no marker: $p\geq0.10$. \textsuperscript{a} indicates the reported window
is later than 2000+ because the indicator's own history starts later
(WIBOR~1Y: 2001-01; Polish CDS: 2008-12; WIRON: 2019-01; Bitcoin:
2010-07) -- see the notes to Appendix Table~\ref{tab:all_correlations}.
Polish CDS's primary source (LSEG/Refinitiv) has several months with no
usable daily quote; these are filled with month-on-month average quotes
from the independent source \citep{su2025dynamic} and from Santander
Bank Polska daily research notes, both documented in the replication
files.
The four window labels used throughout this paper (2000+, 2008+, 2014+,
2020+) are pre-specified and fixed across all indicators and models;
2001+ is not among them, so
WIBOR~1Y (native start 2001-01) is reported at the first of these four
windows that does not precede its own history, 2008+, rather than at a
window matching its exact start date. For the same reason, the reported
$n$ for a given window is not simply its calendar span in months: Polish
CDS's own raw levels start one month later than its nominal 2008+ window
(2008-12, not 2008-01), and the month-on-month \% change transform used
throughout this paper consumes one further observation at the start of
any series (a change needs two consecutive levels); WIRON's own native
history likewise starts only from 2019-01, narrower than its nominal
2020+ window's calendar span would otherwise suggest. The exact
month-by-month coverage underlying every indicator is in the replication
files.
\end{flushleft}
\end{table}

In month-on-month changes the EPU Index PL correlates with the global and the US
EPU index at $+0.248$ and $+0.258$ respectively  (see Table~\ref{tab:epu_correlations}),
both significant at the 1\% level. The association is firm in sign and in
significance but moderate in size: the Polish index moves with the international benchmarks without tracking them. The modest correlations indicate that EPU Index PL contains substantial variation not shared with international benchmarks.

The co-movement with the policy-uncertainty benchmarks is not, however, markedly different in magnitude from that observed for financial-volatility measures. V2X ($+0.277$), VIX ($+0.257$) and MOVE ($+0.196$) exhibit correlations broadly comparable to those of the Global EPU ($+0.248$) and US EPU ($+0.258$) indices. At monthly frequency the component the index shares with foreign series is therefore not identified as foreign \emph{policy} uncertainty specifically; these data do not separate it from global risk. The one external benchmark that does not co-move is geopolitical risk: the GPR index, at $+0.066$, carries no significance marker. This finding is consistent with the event evidence in Section~\ref{sec:events_face}, where neither the invasion of Ukraine (2022-02, $z=-0.7$) nor the 2025 Middle East escalation ($z=+0.3$) coincides with a pronounced increase in EPU Index PL. Together, these results suggest that geopolitical risk and economic policy uncertainty capture distinct, though potentially overlapping, dimensions of uncertainty.

Within the domestic block the pattern is one-directional. Sovereign risk and equity prices move with uncertainty--the Polish CDS premium at $+0.211$, WIG at $-0.203$ and WIG20 at $-0.164$, all at the 1\% level--as do the exchange rates, USD/PLN at $+0.148$ (1\% level) and EUR/PLN at $+0.142$ (5\% level), that is, zloty
depreciation. The significant correlations are consistent with the expected association between heightened policy uncertainty, greater sovereign risk, zloty depreciation and weaker equity-market performance. Money-market rates are the exception: WIBOR at the three-month, six-month and one-year tenors is reported at $-0.037$, $-0.040$ and $-0.059$, the ten-year government yield at
$-0.087$ and WIRON at $+0.096$, none of them significant. Within the domestic financial block, the largest absolute contemporaneous correlations are observed for sovereign risk and equity prices, followed by exchange rates, while associations with money-market rates and government bond yields are statistically insignificant.

Two features of the table bound this reading. First, every correlation marked as significant occurs at lag $t$, while every indicator whose reported best lag is
$t\text{-}1$, $t\text{-}2$ or $t\text{-}3$ is insignificant. The strongest statistically significant associations are contemporaneous; the present correlation analysis does not establish predictive relationships. Second, the reported coefficients represent the largest absolute correlations across four examined lags and should therefore be interpreted as exploratory associations rather than confirmatory evidence of lead--lag relationships. The reported significance levels are nominal and do not adjust for lag selection or multiple testing.

\section{Future Extensions of the EPU Index PL}
\label{sec:future_research}

Future development will focus on policy-specific and sectoral measures of uncertainty. The policy classification framework provides a basis for constructing sub-indices covering fiscal, monetary, tax, regulatory and EU-related policies, with separate validation of their classification accuracy.

Sectoral extensions will follow two complementary approaches: constructing industry-specific EPU indices from sector-relevant news coverage and estimating sectoral exposure to policy-specific uncertainty. This distinction separates the measurement of uncertainty within an industry from its sensitivity to broader policy shocks.

The existing daily EPU Index PL will support high-frequency event studies and analyses of financial-market responses. Further work will strengthen daily-frequency validation and examine, using vector autoregressive (VAR) models of the Polish economy and financial system, the transmission of uncertainty shocks to credit, investment, consumption and financial stability, including their potential predictive value beyond global uncertainty indicators.

\section*{Conclusions}
\addcontentsline{toc}{section}{Conclusions}

This paper introduces the Economic Policy Uncertainty Index for Poland (EPU Index PL), constructed at daily frequency and aggregated into a continuous monthly series covering January 2000 to June 2026. Based on six Polish newspapers and the Baker--Bloom--Davis methodology, it provides a systematically constructed and validated measure of economic policy uncertainty in Poland, extending the internationally comparable EPU framework to the largest economy in Central and Eastern Europe and the sixth-largest in the European Union.

A central methodological contribution is the construction of a consistent historical series from an unbalanced newspaper panel. We demonstrate that changes in newspaper coverage can generate artificial level shifts and that multiplicative chain-splicing preserves non-negativity while better retaining historical variation, particularly during the 2008--2009 Global Financial Crisis. Human validation yields 70.2\% classification precision and 87.5\% pooled inter-annotator agreement. Historical and newspaper-level robustness checks provide further evidence of the index's reliability across periods and editorial orientations.

A central substantive finding is that EPU Index PL captures a predominantly domestic dimension of economic policy uncertainty rather than simply mirroring international uncertainty. This distinction concerns both the object of measurement and the observed variation in the index. By construction, EPU Index PL measures uncertainty surrounding economic policy affecting Poland, including uncertainty arising from domestic decisions and from the domestic policy implications of international developments. Empirically, pronounced responses to selected domestic policy episodes, together with relatively modest correlations with US and Global EPU, support the interpretation that fluctuations in the Polish index are primarily domestically driven. This does not imply insulation from external shocks: international events may contribute to Polish economic policy uncertainty when they create uncertainty about domestic policy responses. The evidence is consistent with this interpretation, although it does not quantitatively decompose domestic and external sources of variation.

The external validation further demonstrates economically interpretable relationships with financial indicators. Increases in EPU Index PL coincide with higher sovereign risk premiums, zloty depreciation and weaker equity-market performance, whereas money-market rates show no statistically significant association. These findings are consistent with the expected relationship between policy uncertainty and financial-market conditions, although the correlations do not establish causality. As a transparent, regularly updated and internationally comparable measure, EPU Index PL provides a foundation for economic research, policymaking, forecasting, stress-testing and financial risk analysis, including future investigations into how domestically generated and externally transmitted uncertainty affect credit, investment and the broader Polish economy.

%% ---------- References ------------------------------------------------
\clearpage
\phantomsection
\addcontentsline{toc}{section}{References}
\bibliography{main}

%% ---------- Appendix --------------------------------------------------
\clearpage
\appendix

%% Tables, figures and equations in the appendices are numbered per appendix (A.1, B.1, ...)
\makeatletter
\@addtoreset{table}{section}\@addtoreset{figure}{section}\@addtoreset{equation}{section}
\makeatother
\renewcommand{\thetable}{\thesection.\arabic{table}}
\renewcommand{\thefigure}{\thesection.\arabic{figure}}
\renewcommand{\theequation}{\thesection.\arabic{equation}}
\renewcommand{\theHtable}{app.\thesection.\arabic{table}}
\renewcommand{\theHfigure}{app.\thesection.\arabic{figure}}
\renewcommand{\theHequation}{app.\thesection.\arabic{equation}}

\section{The E/P/U Framework}
\label{app:dictionary}

\begin{table}[!htbp]
\centering
\caption{Term-set differences across dictionary versions (only items that change; concepts present in every version are omitted)}
\label{tab:dictionary_diff}
\small
\renewcommand{\arraystretch}{1.2}
\begin{tabular}{p{0.36\textwidth} cccccc}
\toprule
\textbf{Item} & \textbf{v1} & \textbf{v2} & \textbf{v3} & \textbf{v4.1/v4.2} & \textbf{v5} & \textbf{v5.1} \\
\midrule
\multicolumn{7}{l}{\emph{E: Economy}} \\
Macro aggregates (PKB, inflacj*, ceny, popyt, podaż)      & -- & \checkmark & \checkmark & -- & -- & -- \\
Business cycle (koniunktur*)                              & -- & \checkmark & \checkmark & \checkmark & \checkmark & \checkmark \\
Macro-/micro-economics (makroekonomi*, mikroekonomi*)     & -- & -- & -- & -- & \checkmark & \checkmark \\
\midrule
\multicolumn{7}{l}{\emph{P: Policy}} \\
Spending (wydatk*)                                        & -- & -- & -- & \checkmark & \checkmark & \checkmark \\
Deficit (deficyt*)                                        & -- & -- & \checkmark & \checkmark & \checkmark & \checkmark \\
Law, general (praw*)                                      & -- & -- & \checkmark & -- & -- & -- \\
Ordinance (rozporządz*)                                   & -- & -- & \checkmark & -- & \checkmark & \checkmark \\
European Union (UE, Bruksel*, dyrektyw*, Komisj* Europ.)  & -- & \checkmark & \checkmark & -- & \checkmark & \checkmark \\
Courts (TSUE, Sąd Najwyższy)                              & -- & \checkmark & \checkmark & -- & -- & -- \\
Head of state (prezyden*)                                 & -- & -- & -- & \checkmark & \checkmark & \checkmark \\
Speaker of parliament (marszał*)                          & -- & -- & -- & -- & \checkmark & \checkmark \\
European institutions, abbreviated (PE, europarlament*, EBC, KE) & -- & -- & -- & -- & -- & \checkmark \\
\midrule
\multicolumn{7}{l}{\emph{U: Uncertainty}} \\
Risk/crisis cluster (kryzys, ryzyko, zagrożeni*, recesj*, spadek, \ldots) & \checkmark & -- & -- & -- & -- & -- \\
Extended synonyms (nieprzewidywal*, niejasn*, niejednoznacz*, wątpliw*)   & $\sim$ & \checkmark & \checkmark & -- & -- & -- \\
Lack of certainty (brak* pewności)                        & -- & -- & -- & -- & \checkmark & \checkmark \\
\midrule
\multicolumn{7}{l}{\emph{Implementation}} \\
Matching                                                  & forms & forms & stems & stems & stems & stems \\
Exclusion rules                                           & -- & -- & -- & \checkmark & -- & -- \\
\bottomrule
\end{tabular}

\vspace{3pt}
\begin{flushleft}
\footnotesize
\textit{Notes:} \checkmark~= present, -- = absent. Only terms whose presence changes across versions are listed. Concepts present in \emph{every} version are omitted for brevity: the core economy stem, and, under Policy, budget (budżet*), tax (podatek/podatk*), regulation (regulac*), legislation (ustaw*/legislac*), parliament (sejm*/sena*/parlament*), government (rząd*), minister (minist*), and
the central bank (NBP/RPP and the ``Narodowy Bank Polski'' / ``Rada Polityki Pieniężnej'' phrases); for these, only the matching form changed--fully inflected forms in v1--v2, prefix wildcards from v3 onward. In the Uncertainty row, $\sim$ denotes that v1 contained only \emph{nieprzewidywal*} and \emph{niejasn*} from the extended-synonym group. The v4.1/v4.2 column reports a single term list because v4.1 and v4.2 share an identical dictionary (see Table~\ref{tab:dictionary_versions}); they differ only in the implementation of the exclusion-rule layer, which is not reproduced here. The v5/v5.1 term lists are provided as online replication files; the superseded v1--v4.2 Boolean queries and the v4.1/v4.2 exclusion rules are available from the authors on request.
\end{flushleft}
\end{table}

\section{Human Audit and Precision Assessment}
\label{app:human_audit}

This appendix reproduces the annotation guidelines used in the manual validation of the EPU Index PL, together with the detailed breakdowns of the audit results and the inter-annotator agreement matrices. The sample design, the adjudication procedure and the headline results are reported in Section~\ref{sec:human_audit}.

\subsection{Annotation guidelines for the EPU Index PL}
\label{app:annotation_guidelines}

\subsubsection{Purpose and scope}

These guidelines serve both as the training material for annotators and as a reference tool during coding. The annotation task is to determine, by reading
the article, whether it discusses economic uncertainty and whether that uncertainty relates to economic policy. Annotators code article content
directly. They do not evaluate the dictionaries, do not evaluate individual keywords, and are not informed of how the automated filter classified the
article. An article may warrant a positive human coding for reasons entirely different from those that triggered the automated classification, and vice
versa; the two judgments are made independently by construction.

\subsubsection{Coding sequence}

Coding proceeds in two steps, and economic policy uncertainty is treated as a
subset of economic uncertainty. An article cannot be coded EPU$=$1 and EU$=$0.

\begin{enumerate}
    \item \textbf{EU.} Is the article about economic uncertainty, even if only     to a limited extent? If yes, set EU$=$1; otherwise EU$=$0.

    \item \textbf{EPU.} If EU$=$1: does the article discuss policy-related     aspects of that economic uncertainty, even if only briefly? If yes, set EPU$=$1; otherwise EPU$=$0.
\end{enumerate}

The article need not be mainly about economic policy uncertainty, and need not contain extensive remarks on the subject, to warrant EPU$=$1. However, if the article discusses economic uncertainty in one place and policy in another, and never connects the two, it does not warrant EPU$=$1.

Grounds for EPU$=$1 include remarks about: uncertainty over who makes or will make policy decisions with economic consequences; current, past or prospective uncertainty over what policy actions will be undertaken; uncertainty regarding the economic effects of policy actions, past, present or future; economic uncertainty induced by policy \emph{inaction}; and economic uncertainty related to policy developments motivated by non-economic considerations, such as
national security.

\subsubsection{Coding fields}

For each article the annotator records the fields listed in
Table~\ref{tab:coding_fields}, using the policy categories in
Table~\ref{tab:policy_categories}.

\begin{table}[!htbp]
\centering
\caption{Coding fields}
\label{tab:coding_fields}
\footnotesize
\begin{tabular}{p{3.1cm}p{11.4cm}}
\toprule
\textbf{Field} & \textbf{Definition} \\
\midrule
\texttt{EU} & 1 if the article is about economic uncertainty, even briefly. \\
\texttt{EPU} & 1 if the article discusses policy-related aspects of economic
uncertainty, even briefly. Requires EU$=$1. \\
\texttt{category\_code} & For EPU$=$1: policy category, all that apply, from
Table~\ref{tab:policy_categories}. \\
\texttt{policy\_term} & For EPU$=$1: the policy term(s) as used in the article
itself, in the article's own wording. Free text. \\
\texttt{p\_terms} & Used only when \texttt{category\_code} is (q): a summary
description of the type of policy uncertainty discussed. \\
\texttt{who} & For EPU$=$1: 1 if the article discusses uncertainty about the
outcome of elections or other contests for political power. \\
\texttt{actions} & For EPU$=$1: 1 if the article discusses uncertainty about
what actions political leaders or officials will take, or when. \\
\texttt{effects} & For EPU$=$1: 1 if the article discusses uncertainty about the
effects of policy decisions, or attributes economic effects to a change in
policy-related uncertainty. Not mutually exclusive with the two preceding
fields. \\
\texttt{down} & 1 if the relevant passages mainly discuss actual or prospective
\emph{declines} in uncertainty. Evaluated on the EPU passages for EPU$=$1 and on
the EU passages for EPU$=$0. \\
\texttt{mention\_foreign} & 1 if the relevant passages mention developments,
concerns or consequences related to foreign actors or their actions. \\
\texttt{mainly\_foreign} & 1 if those passages \emph{mainly} concern foreign
actors. A stricter test: implies \texttt{mention\_foreign}$=$1. \\
\texttt{foreign\_list} & Free text: the foreign countries, regions or
authorities mentioned in connection with uncertainty (e.g.\ Germany, European
Commission, ECB, Ukraine). \\
\texttt{confidence} & 3 --- highly confident, straightforward; 2 --- requires
extra care but confident; 1 --- hard call, on which reasonable annotators could
differ even after careful reading. \\
\texttt{fp\_reason} & Recorded only when EPU$=$0: the error category from the
taxonomy below. \\
\bottomrule
\end{tabular}
\end{table}

\begin{table}[!htbp]
\centering
\caption{Policy categories}
\label{tab:policy_categories}
\footnotesize
\begin{tabular}{p{7.1cm}p{7.1cm}}
\toprule
(a) Monetary policy &
(i) Energy and environmental regulation, natural resources, commodities \\
(b) Fiscal policy (spending and revenue) &
(j) Entitlement programs, social transfers, welfare \\
(c) Taxation &
(k) Financial regulation, banking, capital markets \\
(d) Government spending &
(l) Political conflict and leadership change \\
(e) Labor market regulation &
(m) Sovereign debt, exchange rate policy, reserves \\
(f) Legal and judicial policy &
(n) Trade policy \\
(g) Competition and antitrust policy &
(o) EU policy, EU funds and EU-level regulation \\
(h) Health care programs and regulation &
(p) National security and defense \\
\midrule
\multicolumn{2}{l}{(q) Other policy matters not covered by (a)--(p) --- specify
in \texttt{p\_terms}} \\
\bottomrule
\end{tabular}

\vspace{3pt}
\begin{flushleft}
\footnotesize
\textit{Notes:} The list follows BBD with one addition specific to the Polish
setting: category (o), covering EU-level policy, the Recovery and Resilience
Facility and cohesion funds, which in Polish reporting constitutes a distinct
and quantitatively important source of policy uncertainty and corresponds to the
planned EU-policy sub-index. Categories
are not mutually exclusive.
\end{flushleft}
\end{table}

\subsubsection{Decision rules}

The following rules govern the recurring judgments in the coding task. Each is
stated as a rule with a short illustration in the Polish media context.

\paragraph{R1. Brief mentions count.}
An article need not be mainly about economic policy uncertainty; a single
passage is sufficient. An article about a cultural event that closes with a
remark that demand is uncertain because of a proposed VAT increase is coded
EU$=$1 and EPU$=$1.

\paragraph{R2. Past uncertainty counts.}
Any mention of economic policy uncertainty --- past, present or prospective ---
warrants EPU$=$1. A retrospective analysis arguing that fiscal-policy
uncertainty in an earlier decade deepened a downturn is coded EPU$=$1.

\paragraph{R3. Foreign uncertainty counts.}
Uncertainty about foreign economic policy, or economic uncertainty arising from
the policies of foreign governments and international institutions, warrants
EPU$=$1 regardless of the country concerned. Uncertainty about German fiscal
policy, ECB decisions, EU regulation, US tariffs or Ukrainian reconstruction
policy all qualify. Record the countries or institutions in
\texttt{foreign\_list} and set \texttt{mention\_foreign}$=$1.

\paragraph{R4. No economic vocabulary is required.}
A passage may describe policy-related economic uncertainty without containing
any word such as \textit{gospodarka}, \textit{ekonomiczny} or
\textit{finansowy}. A statement that the future of a permitting regime for a
given industry remains unclear, and that firms are consequently postponing
expansion, warrants EPU$=$1 even though no economic term appears.

\paragraph{R5. A policy response to uncertainty is not by itself EPU.}
A statement that the government cut rates in response to growing economic
uncertainty is not sufficient for EPU$=$1. It becomes EPU$=$1 if the article
also indicates that policy is a source of uncertainty, or that the effects of
the policy are themselves uncertain --- for example, that the measure may raise
uncertainty by widening the deficit, or that it is undecided whether the measure
will be adopted.

\paragraph{R6. Co-occurrence without a link is not EPU.}
An article may discuss both economic uncertainty and government policy without indicating any relationship between them. If the article states, implies or suggests that economic uncertainty is affected by policy actions, inaction, choices or consequences, code EPU$=$1. If the two are presented as unrelated
influences, code EPU$=$0.

\paragraph{R7. A policy proposal without uncertainty is not EPU.}
A description of a new or proposed policy does not by itself warrant EPU$=$1.
The article must state, imply or suggest uncertainty about which policy will be
adopted, whether or when it will be adopted, or what its economic effects will
be. A report that the government has submitted a budget with a given spending
increase is not sufficient; the addition that the parliamentary majority is
contested and the plan may be derailed makes it EPU$=$1.

\paragraph{R8. National security requires economic ramifications.}
Uncertainty related to military action or national security warrants EPU$=$1
only where the article refers to economic consequences --- for example,
uncertainty about the course of the war in Ukraine affecting energy prices,
supply chains or defense procurement. Uncertainty about the prospects for a
ceasefire, with no economic reference, does not.

\paragraph{R9. Judicial and legal matters.}
Three cases are distinguished. If the article merely notes that a case is
unresolved, with no economic ramification, code EU$=$0 and EPU$=$0. If the
economic uncertainty turns on the resolution of a purely factual dispute ---
whether a particular firm caused a particular loss --- code EU$=$1 and EPU$=$0.
If the article invokes economic uncertainty about \emph{whether and how the law
applies} --- for instance, uncertainty over the interpretation of rulings
affecting an entire class of contracts, or over the constitutional status of a
statute with fiscal consequences --- code EPU$=$1, category (f).

\paragraph{R10. Uncertain public revenues.}
Remarks about uncertain tax or budget revenues warrant EPU$=$1 where the article
attributes the uncertainty partly or wholly to uncertainty over policy choices.
They do not warrant EPU$=$1 where the uncertainty is attributed entirely to
economic conditions and those conditions are not themselves linked to policy.

\paragraph{R11. Accuracy of economic statistics.}
Remarks on the reliability of economic data warrant EU$=$1 when presented as an
aspect of an uncertain economic outlook, and do not when presented as an
unrelated methodological aside.

\paragraph{R12. Exclusion rules are not the annotator's concern.}
In dictionary versions v4.1 and v4.2, certain stems (for example \textit{prezydent*}, \textit{budżet*}, \textit{regulac*}) were subject to phrase-level exclusion rules at the pipeline stage (not part of the published pipeline; available from the authors on request). These rules operated on individual term matches and were not part of the annotation task. The published v5/v5.1 pipeline does not apply exclusion rules (see Table~\ref{tab:dictionary_diff}, Appendix~\ref{app:dictionary}): both are a direct E$\wedge$P$\wedge$U match with no post-hoc exclusion filtering. Annotators code the meaning of the article regardless of dictionary version or which terms matched. The audit nevertheless informs the term-set design indirectly, through the distribution of error categories in Table~\ref{tab:false_positive_taxonomy}.

\subsubsection{Taxonomy of false positives}
\label{app:fp_taxonomy}

Where the final coding is EPU$=$0, the annotator records the reason: (1) uncertainty unrelated to economic policy; (2) policy-related terms referring to
non-economic domains; (3) economic terms used in a non-policy context; (4) institutional terms referring to actors or events unrelated to economic policy
decisions; (5) economic, policy and uncertainty elements all present but not connected to one another, in the sense of rule R6. This taxonomy provides the diagnostic input for the subsequent revision of the term sets.

\subsubsection{Worked examples}
\label{app:annotation_examples}

Table~\ref{tab:audit_examples} presents worked examples in the four classes used in annotator training. True positives and false positives are drawn from the audit universe; false negatives are drawn from the near-miss cell of the audit sample itself (articles satisfying E$\wedge$U but no Policy term). 

\begin{table}[!htbp]
\centering
\caption{Worked examples of annotation decisions}
\label{tab:audit_examples}
\footnotesize
\begin{tabular}{p{2.4cm}p{7.2cm}p{4.6cm}}
\toprule
\textbf{Class} & \textbf{Article content} & \textbf{Coding and reason} \\
\midrule
True positive 1 &
\textit{``Kilka miliardów złotych ze swojego zysku może wpłacić do przyszłorocznego budżetu Narodowy Bank Polski. Wpłata będzie możliwa dzięki dokonanej w minionym tygodniu zmianie zasad rachunkowości banku centralnego.'' (\emph{Gazeta Polska Codziennie}, 2019-11-13; uid A0004)} &
EPU$=$1. The article discusses the central bank's accounting-rule change and its planned transfer to the state budget. Category: \textit{(a) Monetary policy}. \\
True positive 2 &
\textit{``Ministrowie handlu krajów UE stwierdzili na nieformalnej naradzie w Brugii (Belgia), że perspektywy uruchomienia nowej rundy negocjacji WTO są nadal niepewne.'' (\emph{Rzeczpospolita}, 2001-09-08; uid A0002)} &
EPU$=$1. The article discusses EU trade ministers' uncertainty over the WTO negotiating round. Category: \textit{(i) Energy and environmental regulation; (a) Monetary policy}. \\
False positive 1 &
\textit{``Ostatnia pandemia potężnie wpłynęła na psychikę i zdrowie społeczeństw oraz na polityki i gospodarki państw. Ekonomiści zadają sobie jednak pytania o to, jak na tę długotrwałą chorobę zareagowali inwestorzy działający na rynkach finansowych [...]'' (\emph{Dziennik Gazeta Prawna}, 2023-06-02; uid A0008)} &
EPU$=$0. Automated filter positive; the article discusses investor psychology and market behavior during the pandemic in general terms, without connecting the uncertainty to a specific economic-policy decision. Error category \textit{(1) uncertainty unrelated to economic policy}. \\
False positive 2 &
\textit{``Firmy, które postawiły na organiczny rozwój, wychodzą z założenia, że ich model biznesu umożliwi szybki wzrost bez przejmowania gotowej działalności.'' (\emph{Parkiet}, 2015-03-30; uid A0010)} &
EPU$=$0. The article is a business-strategy discussion of organic growth versus acquisitions, with no connection to economic-policy uncertainty. Error category \textit{(1) uncertainty unrelated to economic policy}. \\
False negative 1 &
\textit{``Mimo przesunięcia terminu składania ofert na opracowanie elektrobusów w ciągu pięciu lat do samorządów może trafić nawet 1000 takich maszyn, które powstaną w Polsce. Termin składania wniosków o dopuszczenie do postępowania na opracowanie i dostawę pojazdów niskoemisyjnych został wydłużony do 3 kwietnia.'' (\emph{Dziennik Gazeta Prawna}, 2018-03-06; uid A0026)} &
EPU$=$1 by human reading; automated filter negative because the article contains no term from the P set, even though the delayed public-procurement deadline it describes is a policy-driven uncertainty. Category: \textit{(d) Government spending; (q) Other policy matters}. \\
False negative 2 &
\textit{``Dla biznesu najważniejsze zagrożenia to wojna w Ukrainie i inflacja, a wyzwania to transformacja technologiczna i energetyczna, cyberbezpieczeństwo oraz walka o talenty.'' (\emph{Puls Biznesu}, 2023-02-07; uid A0027)} &
EPU$=$1 by human reading; automated filter negative because the article contains no term from the P set. Category: \textit{(i) Energy and environmental regulation; (a) Monetary policy}. \\
Hard call 1 &
\textit{``Rozmowa z Piotrem Kuczyńskim, analitykiem rynku finansowego [...] O kogo powinien dbać bankier? O swoich klientów, czyli klasę średnią.'' (\emph{Gazeta Wyborcza}, 2013-07-04; uid A0036). Policy terms discussed: państwo opiekuńcze, prywatyzacja, deregulacja zawodów, niskie podatki, elastyczny rynek pracy.} &
Recommended EPU$=$1, confidence 1. The article is an opinion interview ranging broadly over welfare-state, privatization and labor-market philosophy rather than a specific policy decision, so reasonable annotators could code the E--P--U link as present but diffuse or as insufficiently connected (rule R6). Category: \textit{(c) Taxation}. \\
Hard call 2 &
\textit{``Trwającą od roku poprawę notowań trzeba też rozpatrywać w kontekście nowych zagrożeń [...] zmiana sposobu wyliczania norm formaldehydów w meblach sprzedawanych w Niemczech.'' (\emph{Puls Biznesu}, 2020-03-02; uid A0037)} &
Recommended EPU$=$1, confidence 1. EPU rests on a short passage about the uncertain effect of German formaldehyde-emission norms on the furniture maker's sales; the regulatory link is real but narrow and buried within an otherwise financial article. Category: \textit{(i) Energy and environmental regulation}. \\
\bottomrule
\end{tabular}

\vspace{3pt}
\begin{flushleft}
\footnotesize
\textit{Notes:} Hard calls are included in the training material to calibrate
annotators on the boundary of the definition rather than only on clear cases.
Each example reproduces enough of the article to make the coding decision
verifiable by a reader.
\end{flushleft}
\end{table}

\subsection{Audit results: detailed breakdown}
\label{app:audit_breakdown}

Table~\ref{tab:human_audit} summarizes the annotation and validation procedure.

\begin{table}[!htbp]
\centering
\caption{Human annotation and validation procedure}
\label{tab:human_audit}
\small
\renewcommand{\arraystretch}{1.2}
\begin{tabular}{p{0.16\textwidth} p{0.30\textwidth} p{0.22\textwidth} p{0.20\textwidth}}
\toprule
\textbf{Component} & \textbf{Sample} & \textbf{Purpose} & \textbf{Metric} \\
\midrule

\textbf{Training / calibration}
& 60 training articles with a planned class mix (true positives, false positives, false negatives, hard calls), coded by all three annotators; disjoint from the audit sample.
& Calibrate annotators on boundary cases before the main audit.
& Not reported (training only) \\
\midrule

\textbf{Main human audit}
& Two-stage stratified sample of 600 articles from the E$\wedge$U universe: 480 flagged (E$\wedge$P$\wedge$U) $+$ 120 near-miss (E$\wedge$U$\wedge\neg$P), by newspaper $\times$ month (20 months per title; 4$+$1 per month), fixed seed; 200 primary articles per annotator.
& Validate the E$\wedge$P$\wedge$U rule in both directions: precision on the flagged cell and gross error over E$\wedge$U.
& Confusion matrix (EPUH$\times$EPUC); precision (design-based: Wilson CI; volume-weighted: normal-approximation CI); FP/FN rate; gross error \\
\midrule

\textbf{Inter-annotator agreement}
& Overlap of 120 articles (three annotator pairs $\times$ 40), drawn from the audit sample.
& Consistency of independent codings beyond chance.
& Observed agreement; Cohen's $\kappa$ per pair (3 values); pooled Fleiss' $\kappa$, PABAK, Gwet's $AC_1$; 95\% CI and $N$; disagreement by confidence \\
\midrule

\textbf{Error analysis and adjudication}
& Disagreements and borderline cases from the overlap sample.
& Resolve disagreements; classify false positives; refine coding rules.
& Adjudicated labels (third annotator, who did not code the article) feeding the results above; false-positive taxonomy (5 categories) \\

\bottomrule
\end{tabular}

\vspace{3pt}
\begin{flushleft}
\footnotesize
\textit{Notes:} Annotators code article content blind to the automated classification (no algorithm label, no highlighted keyword matches), see the full article text, and receive flagged and near-miss articles interleaved in randomized order. The audit universe is the E$\wedge$U set under the EPU Index PL dictionary; sampling uses a fixed seed. The main audit identifies both Type~I errors (false positives, flagged cell) and Type~II errors (false negatives, near-miss cell); precision and the gross error rate are computed on post-adjudication labels with population weights.
\end{flushleft}
\end{table}
Tables~\ref{tab:human_audit_by_title} and~\ref{tab:human_audit_by_period} report precision by newspaper and by period, respectively.

\begin{table}[!htbp]
\centering
\caption{Precision by newspaper}
\label{tab:human_audit_by_title}
\small
\begin{tabular}{lrrr}
\toprule
\textbf{Newspaper} & \textbf{Flagged audited} & \textbf{Precision (\%)} & \textbf{95\% CI} \\
\midrule
Rzeczpospolita           & 80 & 66.3 & 55.4--75.7 \\
Parkiet                  & 80 & 87.5 & 78.5--93.1 \\
Dziennik Gazeta Prawna   & 80 & 75.0 & 64.5--83.2 \\
Gazeta Wyborcza          & 80 & 40.0 & 30.0--51.0 \\
Gazeta Polska Codziennie & 80 & 78.8 & 68.6--86.3 \\
Puls Biznesu             & 80 & 73.8 & 63.2--82.1 \\
\midrule
Full sample              & 480 & 70.2 & 66.0--74.1 \\
\bottomrule
\end{tabular}
\vspace{3pt}
\begin{flushleft}
\footnotesize
\textit{Notes:} Precision on the flagged cell only ($n=80$ per title). Wilson intervals. Equal allocation across titles; a volume-weighted aggregate is in Table~\ref{tab:human_audit_results}. Following the 2026-10-20 redraw of 20 out-of-window articles (Section~\ref{sec:adjudication}), all 20 replaced articles are Rzeczpospolita; no other newspaper row is affected.
\end{flushleft}
\end{table}

\begin{table}[!htbp]
\centering
\caption{Precision by period}
\label{tab:human_audit_by_period}
\small
\begin{tabular}{lrrrr}
\toprule
\textbf{Period} & \textbf{Titles available} & \textbf{Audited} &
\textbf{Precision (\%)} & \textbf{95\% CI} \\
\midrule
2000--2007 & 2 & 64  & 79.7 & 68.3--87.7 \\
2008--2014 & 5 & 92  & 66.3 & 56.2--75.1 \\
2015--2019 & 6 & 92  & 67.4 & 57.3--76.1 \\
2020--2026 & 6 & 232 & 70.3 & 64.1--75.8 \\
\midrule
Full sample & & 480 & 70.2 & 66.0--74.1 \\
\bottomrule
\end{tabular}

\vspace{3pt}
\begin{flushleft}
\footnotesize
\textit{Notes:} The number of titles available differs across periods because of archival coverage (see Table~\ref{tab:newspaper_history}). The period breakdown is reported to establish that measured precision is not driven by high-volume crisis episodes. Following the 2026-10-20 redraw of 20 out-of-window Rzeczpospolita articles (Section~\ref{sec:adjudication}), all 20 replaced articles fall in the 2000--2007 period; no other period row is affected.
\end{flushleft}
\end{table}

Table~\ref{tab:coding_confidence} reports the distribution of coding confidence in the audit sample.

\begin{table}[!htbp]
\centering
\caption{Distribution of coding confidence}
\label{tab:coding_confidence}
\small
\begin{tabular}{lrr}
\toprule
\textbf{Coding confidence} & \textbf{Articles} & \textbf{Share (\%)} \\
\midrule
3 --- high confidence, straightforward   & 411 & 68.5 \\
2 --- requires care, confident           & 170 & 28.3 \\
1 --- hard call                          & 19  &  3.2 \\
\midrule
Total                                    & 600 & 100.0 \\
\bottomrule
\end{tabular}

\vspace{3pt}
\begin{flushleft}
\footnotesize
\textit{Notes:} Confidence is coded by the annotator whose classification was
retained after adjudication. Level 1 denotes articles on which reasonable
annotators could reach different judgments even after careful reading.
\end{flushleft}
\end{table}

\subsection{Inter-annotator agreement}
\label{app:inter_annotator}

Agreement matrices are reported for the overlap subsample of 120 articles, on the pre-adjudication codings of the binary EPU field. Each of the three
annotator pairs coded 40 articles in common. Summary statistics are reported in
Table~\ref{tab:agreement_stats} in the main text.

Table~\ref{tab:kappa_matrix} reports the pooled agreement matrix and Table~\ref{tab:kappa_matrix_pairs} the matrices for each annotator pair.

\begin{table}[!htbp]
\centering
\caption{Pooled agreement matrix, all pairs}
\label{tab:kappa_matrix}
\small
\begin{tabular}{lccc}
\toprule
& \textbf{Second coder: EPU$=$1} & \textbf{Second coder: EPU$=$0}
& \textbf{Total} \\
\midrule
\textbf{First coder: EPU$=$1}      & 66 & 9 & 75 \\
\textbf{First coder: EPU$=$0} & 6 & 39 & 45 \\
\midrule
\textbf{Total}                       & 72 & 48 & 120 \\
\bottomrule
\end{tabular}

\vspace{3pt}
\begin{flushleft}
\footnotesize
\textit{Notes:} The $(1,1)$ cell counts articles coded EPU$=$1 by both coders and the $(2,2)$ cell those coded EPU$=$0 by both; the
off-diagonal cells are disagreements, resolved by the third annotator under the procedure described in Section~\ref{sec:adjudication}. Observed agreement $=$
87.5\%; Cohen's $\kappa=$~0.74 (95\% CI 0.60--0.85); PABAK $=$~0.75; Gwet's $AC_1=$~0.76. The assignment of coders to the ``first'' and ``second'' positions
within a pair is arbitrary; the matrix is reported in the orientation used for the $\kappa$ computation.
\end{flushleft}
\end{table}

\begin{table}[!htbp]
\centering
\caption{Agreement matrices by annotator pair}
\label{tab:kappa_matrix_pairs}
\small
\begin{tabular}{llcc}
\toprule
\textbf{Pair} & \textbf{First coder} & \textbf{EPU$=$1} &
\textbf{EPU$=$0} \\
\midrule
A1--A2 & EPU$=$1       & 20 & 5 \\
       & EPU$=$0        & 5  & 10 \\
\midrule
A1--A3 & EPU$=$1       & 25 & 3 \\
       & EPU$=$0        & 0  & 12 \\
\midrule
A2--A3 & EPU$=$1       & 21 & 1 \\
       & EPU$=$0        & 1  & 17 \\
\bottomrule
\end{tabular}

\vspace{3pt}
\begin{flushleft}
\footnotesize
\textit{Notes:} Rows correspond to the coding of the first-listed annotator of
each pair, columns to that of the second. $n=40$ per pair. Pairwise $\kappa$
values are reported in Table~\ref{tab:agreement_stats}.
\end{flushleft}
\end{table}

Table~\ref{tab:disagreement_by_confidence} relates disagreements to the coders' confidence levels.

\begin{table}[!htbp]
\centering
\caption{Disagreements by coding confidence}
\label{tab:disagreement_by_confidence}
\small
\begin{tabular}{lrrr}
\toprule
\textbf{Minimum confidence in pair} & \textbf{Articles} &
\textbf{Disagreements} & \textbf{Agreement (\%)} \\
\midrule
3 & 49 & 2 & 95.9 \\
2 & 58 & 9 & 84.5 \\
1 & 13 & 4 & 69.2 \\
\midrule
All & 120 & 15 & 87.5 \\
\bottomrule
\end{tabular}

\vspace{3pt}
\begin{flushleft}
\footnotesize
\textit{Notes:} Articles are grouped by the lower of the two confidence values
assigned within the pair. Concentration of disagreements in the lowest
confidence band indicates that residual disagreement reflects genuinely
ambiguous cases rather than inconsistent application of the guidelines.
\end{flushleft}
\end{table}

\section{Media-Slant Robustness Test}
\label{app:slant}

\subsection{Newspaper classification}

\begin{table}[!htbp]
\centering
\caption{Newspaper classification used in the media-slant robustness test}
\label{tab:slant_classification}
\small
\begin{tabular}{ll}
\toprule
\textbf{Editorial profile} & \textbf{Newspapers} \\
\midrule
Left-leaning & Gazeta Wyborcza \\
Right-leaning & Gazeta Polska Codziennie \\
Business-oriented & Puls Biznesu, Parkiet \\
National economic and policy press & Rzeczpospolita, Dziennik Gazeta Prawna \\
\bottomrule
\end{tabular}
\begin{flushleft}
\footnotesize
\textit{Notes:} Profiles characterize the panel and enter only the tests of
Section~\ref{sec:slant}; they do not enter the construction of the index.
Only GW and GPC are treated as partisan; the remaining titles are not
assigned a political side.
\end{flushleft}
\end{table}

\subsection{Title-by-title correlation matrix}

Table~\ref{tab:slant_correlation} reports the pairwise correlations between
the newspaper-level series (month-on-month changes, common sample). It is the
input to the direct comparison in Section~\ref{sec:slant_direct}; the
liberal--conservative pair is the GW$\times$GPC entry.

\begin{table}[!htbp]
\centering
\caption{Pairwise correlations between newspaper-level EPU series (changes)}
\label{tab:slant_correlation}
\small
\begin{tabular}{lrrrrrr}
\toprule
 & \textbf{GW} & \textbf{GPC} & \textbf{PB} & \textbf{Parkiet} & \textbf{Rz} & \textbf{DGP} \\
\midrule
GW       & 1.000 \\
GPC      & 0.074 & 1.000 \\
PB       & 0.154 & 0.062 & 1.000 \\
Parkiet  & 0.133 & 0.160 & 0.337 & 1.000 \\
Rz       & $-$0.095 & 0.087 & 0.269 & 0.233 & 1.000 \\
DGP      & 0.084 & $-$0.099 & 0.104 & 0.110 & 0.162 & 1.000 \\
\bottomrule
\end{tabular}
\begin{flushleft}
\footnotesize
\textit{Notes:} Pearson correlations of first differences over the common
sample; $n$ per pair reported in the replication files. The matrix is
title-by-title only; correlations of each title with the aggregate are not
reported, being mechanically inflated by the title's own inclusion.
\end{flushleft}
\end{table}

\clearpage
\section{Correlations with uncertainty, financial and macroeconomic indicators}
\label{app:correlations}

\begin{table}[!ht]
\centering
\caption{Correlation results between EPU Index PL and uncertainty, financial and macroeconomic indicators, month-on-month \% changes, 2000+ window (or the earliest window supported by the indicator's own history -- see \textsuperscript{a} notes)}
\label{tab:all_correlations}
\small\renewcommand{\arraystretch}{0.92}

\begin{tabular}{llrrrr}
\toprule
\textbf{Indicator} &
\textbf{Window} &
\textbf{t} &
\textbf{t-1} &
\textbf{t-2} &
\textbf{t-3}
\\
\midrule

\multicolumn{6}{l}{\textbf{Global uncertainty benchmarks}}\\
Global EPU Index & 2000+ & 0.248 & 0.157 & $-$0.042 & $-$0.037 \\
US EPU Index & 2000+ & 0.258 & 0.085 & $-$0.026 & $-$0.057 \\

\midrule
\multicolumn{6}{l}{\textbf{Geopolitical uncertainty}}\\
GPR Index (30-day MA) & 2000+ & 0.066 & 0.052 & $-$0.040 & 0.020 \\

\midrule
\multicolumn{6}{l}{\textbf{Financial volatility}}\\
VIX & 2000+ & 0.257 & 0.016 & $-$0.057 & 0.004 \\
MOVE & 2000+ & 0.196 & $-$0.071 & $-$0.003 & 0.019 \\
V2X & 2000+ & 0.277 & $-$0.015 & $-$0.045 & 0.002 \\

\midrule
\multicolumn{6}{l}{\textbf{Domestic financial conditions}}\\
EUR/PLN & 2000+ & 0.142 & 0.055 & 0.076 & $-$0.060 \\
USD/PLN & 2000+ & 0.148 & 0.108 & $-$0.021 & $-$0.079 \\
GBP/PLN & 2000+ & 0.115 & $-$0.013 & $-$0.072 & $-$0.081 \\
CHF/PLN & 2000+ & 0.185 & 0.063 & 0.037 & $-$0.073 \\

\midrule
\multicolumn{6}{l}{\textbf{Equity markets}}\\
WIG & 2000+ & $-$0.203 & $-$0.022 & $-$0.010 & 0.030 \\
WIG20 & 2000+ & $-$0.164 & $-$0.024 & 0.004 & 0.037 \\
EuroStoxx 50 & 2000+ & $-$0.233 & $-$0.060 & $-$0.036 & 0.025 \\

\midrule
\multicolumn{6}{l}{\textbf{Interest rates and monetary conditions}}\\
WIBOR 1Y\textsuperscript{a} & 2008+ & $-$0.003 & $-$0.059 & $-$0.022 & $-$0.024 \\
WIBOR 6M & 2000+ & 0.017 & $-$0.040 & $-$0.012 & $-$0.006 \\
WIBOR 3M & 2000+ & 0.021 & $-$0.037 & $-$0.010 & $-$0.008 \\
WIRON\textsuperscript{a} & 2020+ & 0.020 & $-$0.060 & 0.003 & 0.096 \\
PL 10Y government yield & 2000+ & 0.072 & $-$0.002 & $-$0.087 & $-$0.062 \\

\midrule
\multicolumn{6}{l}{\textbf{Sovereign risk}}\\
Polish CDS / sovereign spread\textsuperscript{a} & 2008+ & 0.211 & 0.032 & 0.024 & 0.023 \\

\midrule
\multicolumn{6}{l}{\textbf{Safe-haven and risk-sensitive assets}}\\
Gold & 2000+ & 0.030 & 0.006 & 0.006 & $-$0.018 \\
Bitcoin\textsuperscript{a} & 2014+ & $-$0.029 & $-$0.041 & 0.008 & $-$0.071 \\

\bottomrule
\end{tabular}

\vspace{3pt}

\begin{flushleft}
\footnotesize
\textit{Notes:} Pearson correlations between month-on-month \% changes of
the EPU Index PL (\texttt{epu\_strict\_v51\_mult}, multiplicative-splice
vintage) and month-on-month \% changes of each indicator, at each lag
$t/t\text{-}1/t\text{-}2/t\text{-}3$ (EPU leading the indicator by that
many months). Reported for the FIRST of the paper's four
evidence-selected windows (2000+/2008+/2014+/2020+) that the indicator's
own native data history supports; \textsuperscript{a} marks a window
later than 2000+ because the indicator's own history starts later
(WIBOR~1Y: 2001-01; Polish CDS via the LSEG-refreshed series: 2008-12;
WIRON: 2019-01; Bitcoin: 2010-07). Earlier nominal windows are omitted
for these rows rather than repeated, since they would silently collapse to an identical dataset and result.
The main text (see Table~\ref{tab:epu_correlations}) reports the single
best-|r| lag per indicator with significance stars only (no printed
p-values); this table reports every lag for full transparency. Exact
p-values for every cell are provided in the online replication files. The corresponding LEVELS-
transform correlations (main-text Table~\ref{tab:epu_correlations}'s
methodology note explains why these are not the primary evidence) and
all four windows for every indicator are provided in the replication files.
\end{flushleft}

\end{table}

\makebackcover

\end{document}